\documentclass[twocolumn,english,prb,showpacs,amsmath,amssymb,eqsecnum,preprintnumbers]{revtex4}
\usepackage[T1]{fontenc}
\usepackage[latin9]{inputenc}
\usepackage{color}
\usepackage{bm}
\usepackage{amstext}
\usepackage{esint}
\usepackage{soul}
\usepackage{graphicx}
\usepackage{dcolumn}
\usepackage{bigstrut}
\usepackage{rotating}    

\makeatletter
\@ifundefined{textcolor}{}
{%
 \definecolor{BLACK}{gray}{0}
 \definecolor{WHITE}{gray}{1}
 \definecolor{RED}{rgb}{1,0,0}
 \definecolor{GREEN}{rgb}{0,1,0}
 \definecolor{BLUE}{rgb}{0,0,1}
 \definecolor{CYAN}{cmyk}{1,0,0,0}
 \definecolor{blue}{cmyk}{0,1,0,0}
 \definecolor{YELLOW}{cmyk}{0,0,1,0}
}

\def\kF{k_{\text{F}}}

\def\be{\begin{equation}}
\def\ee{\end{equation}}
\def\bea{\begin{eqnarray}}
\def\eea{\end{eqnarray}}
\def\bse{\begin{subequations}}
\def\ese{\end{subequations}}

\def\Tc{T_{\text{c}}}
\usepackage{dcolumn}
\usepackage{bm}
\usepackage{amsfonts}
\draft

\makeatother

\usepackage{babel}
\begin{document}
\preprint{Phys. Rev. B {\bf 91}, 214407 (2015) (corrected version)}

\title{Exponent relations at quantum phase transitions, with applications to metallic quantum ferromagnets}

\author{T. R. Kirkpatrick$^{1}$, D. Belitz$^{2,3}$}

\affiliation{$^{1}$ Institute for Physical Science and Technology,and Department
of Physics, University of Maryland, College Park, MD 20742, USA\\
$^{2}$ Department of Physics and Institute of Theoretical Science,
University of Oregon, Eugene, OR 97403, USA\\
$^{3}$ Materials Science Institute, University of Oregon, Eugene,
OR 97403, USA\\
  }

\date{\today}
\begin{abstract}
Relations between critical exponents, or scaling laws, at both continuous and discontinuous quantum phase transitions 
are derived and discussed. In general there are multiple dynamical exponents at these transitions, which
complicates the scaling description. Some rigorous inequalities are derived, and the conditions needed 
for these inequalities to be equalities are discussed. Scaling laws involving the specific-heat exponents that are
specific to quantum phase transitions are derived and and contrasted with their counterparts at classical phase transitions. 
We also generalize the ideas of Fisher and Berker and others for applying (finite-size) scaling theory near a classical
first-order transition to the quantum case. We then apply and illustrate all of these ideas by using the quantum
ferromagnetic phase transition in metals as an explicit example. This transition is known to have multiple
dynamical scaling exponents, and in general it is discontinuous in clean systems, but continuous in disordered ones.
Furthermore, it displays many experimentally relevant crossover phenomena that can be described in terms of
fixed points, originally discussed by Hertz, that ultimately become unstable asymptotically close to the transition
and give way to the asymptotic fixed points.
These fixed points provide a rich environment for illustrating the general scaling concepts and exponent relations. 
We also discuss the quantum-wing critical point at the tips of the tricritical wings associated with the discontinuous quantum
ferromagnetic transition from a scaling point of view.
\end{abstract}
\maketitle

\section{Introduction}
\label{sec:I}

Classical or thermal phase transitions are special points in parameter space where the free energy
is not analytic as a consequence of strong thermal fluctuations, see, e.g., Refs.~\onlinecite{Stanley_1971,
Wilson_Kogut_1974, Ma_1976, Fisher_1983}. It is customary
to distinguish between second-order or continuous phase transitions, where the
order parameter (OP)\cite{OP_footnote} goes to zero continuously at the transition, and first-order or
discontinuous ones, where the order parameter displays a discontinuity. Classic
examples for the former are the ferromagnetic transition at the Curie temperature
in zero magnetic field, or the liquid-gas transition at the critical point; for the latter,
a ferromagnet below the Curie temperature in a magnetic field, or the liquid-gas
transition below the critical pressure. The nonanalytic free energy translates into
singular behavior of observables. At a second-order transition, this usually takes
the form of power laws, in which case the singularity is characterized by critical 
exponents. Let $m$ be the order parameter (in the case of a ferromagnet, $m$
is the magnetization, in the case of a liquid-gas transition, the density difference
between the phases), $h$ the external field conjugate to the order parameter, 
and $r$ the dimensionless distance from the critical point for $h=0$;\cite{critical_isochore_footnote}
at a thermal phase transition, $r$ is usually chosen to be the distance from the
critical temperature $\Tc$, $r = T/\Tc - 1$. Then the specific heat $C$ behaves as 
$C \propto \vert r\vert^{-\alpha}$, the order parameter $m$ vanishes according
to $m(r\to 0-) \propto (-r)^{\beta}$, the order-parameter susceptibility $\chi$ diverges 
according to $\chi(T) \propto \vert r\vert^{-\gamma}$, etc.\cite{exponents_prime_footnote}
$\alpha$, $\beta$, and $\gamma$ are examples of critical exponents. Underlying all of these singularities is
a diverging length scale, the correlation length $\xi$. It measures the distance
over which order-parameter fluctuations are correlated, and diverges as 
$\xi \propto \vert r\vert^{-\nu}$, which defines the exponent $\nu$. Two other important critical exponents are $\delta$,
which describes the behavior of the order parameter as a function of its conjugate
field $h$ at criticality, $m(r=0,h) \propto h^{1/\delta}$, and the exponent $\eta$,
which describes the decay of the order-parameter correlation function at
criticality, $G(r=0,\vert{\bm x}\vert \to \infty) = \langle m({\bm x})\,m(0)\rangle \propto
1/\vert{\bm x}\vert^{d-2+\eta}$, where $d$ is the spatial dimensionality.

The critical exponents are not all independent. At most classical transitions, it turns
out that specifying two exponents determines all of the others. This was
recognized early on and put in the form of exponent relations (also referred to as
``scaling relations'', or ``scaling laws''; the latter not to be confused with the
homogeneity laws that are often called ``scaling laws'').
For instance, the six exponents $\alpha$, $\beta$, $\gamma$, $\delta$, $\eta$, and
$\nu$ are related by the four relations listed in Eqs.~(\ref{eqs:B.1}). These were
initially derived from scaling assumptions, i.e., generalized homogeneity laws for
various observables near a critical point, and later understood more deeply in
terms of the renormalization group (RG), which allowed for a derivation of the homogeneity
laws. Some of the exponent relations are more robust than others. For instance,
for a $\phi^4$-theory in $d>4$ spatial dimensions, the Widom, Fisher, and
Essam-Fisher scaling relations expressed in Eqs.\ (\ref{eqs:B.1}) remain valid,
whereas the hyperscaling relation, Eq.\ (\ref{eq:B.1d}), fails. This is related to the
notion of dangerous irrelevant variables (DIVs), i.e., coupling constants that flow to zero
under RG transformations but that some observables depend on in a singular way.
As a result, ``strong scaling'', i.e., a simple homogeneity law for the free energy,
breaks down. More generally, there are constraints on the critical exponents that 
take the form of inequalities. They rely on much weaker assumptions than strong 
scaling and in some cases are rigorous. Examples are the Rushbrooke inequality,
Eq.~(\ref{eq:B.5}), and the lower bound for the correlation-length exponent $\nu$
in disordered systems that is discussed in Appendix~\ref{app:C}. 

A quantum phase transition (QPT) occurs by definition at zero temperature, $T=0$,
as a function of some non-thermal control parameter such as pressure, composition,
or an external magnetic field.\cite{Hertz_1976} The critical behavior at $T=0$ is governed by quantum
fluctuations rather than thermal ones. The role of temperature is thus different from that
at a thermal transition, which necessitates the introduction of additional critical
exponents. 
For instance, the order-parameter susceptibility will vary as $\vert r\vert^{-\gamma}$, with
$r$ the dimensionless distance at $T=h=0$ from the critical point, but as $T^{-\gamma_T}$
as a function of temperature for $r\to 0$, with $\gamma_T$ in general different from $\gamma$.  Similarly, 
we need to define exponents $\beta_T$ and $\nu_T$ that describe the behavior of the order-parameter
and the correlation length, respectively, as functions of the temperature in addition to $r$.

The case of the specific heat $c$ is less straightforward since $c$
vanishes at $T=0$ even away from any critical point. Related to this, in the thermodynamic
identity that underlies the Rushbrooke inequality the specific-heat coefficient $\gamma = c/T$
appears. For the purpose of discussing quantum phase transitions it therefore is
sensible to define critical exponents $\bar\alpha$ and ${\bar\alpha}_T$ that describe the
 critical behavior of the specific-heat coefficient according to 
 $\gamma(r,T=0) \propto \vert r\vert^{-{\bar\alpha}}$,
and $\gamma(r=0,T\to 0) \propto T^{-{\bar\alpha}_T}$. For thermal phase transitions, 
${\bar\alpha}$ obviously coincides with the ordinary specific-heat exponent $\alpha$.
For QPTs, we define $\alpha$ and $\alpha_T$ in terms of the susceptibility $\chi_r$ given by the second derivative
of an appropriate free energy with respect to $r$; the physical meaning of this susceptibility
depends on the nature of $r$.
The definition of critical exponents discussed in this paper is summarized in
Appendix \ref{app:A}, and we will refer to the exponents $\alpha$, $\bar\alpha$, $\beta$, etc. as the
$r$-exponents, and to $\alpha_T$, ${\bar\alpha}_T$, $\beta_T$, etc. as the $T$-exponents. 

The concepts of scaling, and exponent relations, also carry over to dynamical critical
phenomena.\cite{Ma_1976, Hohenberg_Halperin_1977} A dynamical critical exponent $z$ is defined by how the critical time scale
$\tau_{\xi}$ scales with the diverging correlation length $\xi$: $\tau_{\xi}\propto \xi^z$. 
In the case of thermal phase transitions, the dynamics are decoupled from the
statics, and the static exponents can be determined independently of $z$. At quantum phase
transitions this is not true since the statics and dynamics are intrinsically coupled. However, as
we will show, at $T=0$ there still are many exponent relations that involve the static exponents only.
A further complication is the fact that multiple critical time scales, and hence multiple
dynamical exponents $z$, which do occur at classical transitions \cite{DeDominicis_Peliti_1978} but are not common,
are the rule rather than the exception at quantum phase transitions.

Interestingly, the same concepts can also be applied to first-order phase transitions.
Fisher and Berker \cite{Fisher_Berker_1982} have developed a scaling description
of classical first-order transitions and have shown that they can be understood as a limiting
case of second-order transitions or critical points. 

The purpose of this paper is to thoroughly discuss these concepts in the context of quantum phase transitions.
While scaling concepts have been used since the notion of quantum phase transitions was
invented, the issue of multiple dynamical exponents and its interplay with dangerous
irrelevant variables has never been systematically discussed. Similarly, the importance of
exponent relations has not been stressed in this context, and even the distinction between 
$r$-exponents and $T$-exponents is not always clearly made. In the first part of the paper
we discuss scaling concepts for quantum phase transitions in general, with an emphasis
on which aspects can be taken over from classical phase-transition theory and which ones
require modifications or new ideas. In particular, we show that the specific-heat coefficient
shows scaling behavior that sets it apart from other observables and is not in direct analogy
to the classical case, and we derive exponent relations that involve the specific-heat
exponents. Exponent relations have proven very useful in the classical context, since they 
provide stringent checks for whether experiments are actually in the asymptotic
region. They are expected to be similarly useful for QPTs. We also generalized the scaling
theory for thermal first-order transitions by Fisher and Berker\cite{Fisher_Berker_1982} to the 
quantum case. In the second part of the paper we apply all of these concepts to the problem
of the quantum ferromagnetic transition in metals. This transition is known to be generically
first order in clean systems, and second order in disordered ones, and it has multiple dynamic
critical exponents, which allows for an illustration of all of the ideas developed in the first part.
It also displays many crossover phenomena the understanding of which is crucial for the interpretation
of experiments. These crossovers can be phrased in terms of fixed points, originally discussed
by Hertz,\cite{Hertz_1976} that become unstable asymptotically close to the transition. The
experimental relevance of Hertz's fixed point in disordered systems in particular is discussed
here for the first time. 

This paper is organized as follows. In Sec.\ \ref{sec:II} we derive exponent relations for the quantum case, 
both for the usual exponents and for those unique to
QPTs, and we generalize the notion of scaling at a first-order transition to the quantum case. 
In Sec.\ \ref{sec:III} we apply our results to the problem of the quantum
ferromagnetic transition in metals, and in Sec.\ \ref{sec:IV} we conclude with a general discussion.
Definitions of common critical exponents are given in Appendix \ref{app:A}, and various
classical exponent relations are recalled in Appendix \ref{app:B}. 
Appendix \ref{app:C} summarizes the Harris criterion and its generalization for the
correlation-length exponent in systems with quenched disorder.

\section{Scaling at quantum phase transitions}
\label{sec:II}

\subsection{General concepts for quantum phase transitions}
\label{subsec:II.A}

In this section we discuss some very general points related to the role of dangerous
irrelevant variables and multiple time scales in the context of quantum phase transitions. 

\subsubsection{Dangerous irrelevant variables and dynamical scaling}
\label{subsubsec:II.A.1}

The classical exponent relations recalled in Appendix \ref{app:B} fall into two distinct classes:
The Widom, Essam-Fisher, and Fisher relations hold quite generally, including the case of
systems above an upper critical dimensionality. By contrast, the hyperscaling relation, 
Eq.\ (\ref{eq:B.1d}), holds only below an upper critical dimension, when no dangerous
irrelevant variables (DIVs) are present. At quantum critical points the role of
DIVs is more complex than in the classical case. A DIV can affect the temperature
scaling of an observable, and it can do so with or without also affecting the static
scaling. We will refer to the assumptions underlying scaling that can or cannot be
altered by DIVs as ``strong'' and ``weak'' scaling assumptions, respectively, and to the
resulting scaling behavior as ``weak scaling'' and ``strong scaling''. We will further
distinguish between weak and strong static and dynamic scaling as appropriate. With this
nomenclature, hyperscaling in the usual sense is a special case of strong static scaling. (Note
that weak scaling is more robust than strong scaling.) 
The DIV concept is more important for
quantum phase transitions than for classical ones, since the latter are more likely to be above their
upper critical dimension and strong scaling is more often violated. 

To illustrate these points, consider an observable ${\cal O}$. Let $[{\cal O}]$ be the scale dimension
of ${\cal O}$ as determined by power counting within the framework of an appropriate field theory, 
and define a critical exponent $\omega$ by the critical behavior of ${\cal O}$ as a function of the
dimensionless distance from criticality at $T=0$, which we denote by $r$: ${\cal O}(r,T=0) \propto \vert r\vert^{\omega}$.
If strong static scaling holds, ${\cal O}$ obeys a homogeneity law\cite{scaling_parts_footnote}
\bse
\label{eqs:2.1}
\be
{\cal O}(r,T=0) = b^{-[{\cal O}]}\,\Phi_{\cal O}(r\,b^{1/\nu}, T=0)\ .
\label{eq:2.1a}
\ee
Here $\nu$ is the correlation length exponent, and $\Phi_{\cal O}$ is a scaling function. Strong
scaling thus leads to 
\be
\omega = \nu [{\cal O}]\ . 
\label{eq:2.1b}
\ee
\ese
This does not remain true if there is an irrelevant
variable that is dangerous with respect to the $r$-dependence of ${\cal O}$. This changes the
value of $\omega$, but usually it does not change the fact that ${\cal O}$ obeys a homogeneity
law. Weak static scaling thus still holds in general,\cite{weak_scaling_footnote} as expressed by
\be
{\cal O}(r,T=0) = b^{-\omega/\nu}\,\Phi_{\cal O}(r\,b^{1/\nu}, T=0)\ .
\label{eq:2.2}
\ee
The value of $\omega$ is now affected by the pertinent DIV and can no longer be determined
by scaling arguments alone, even if $[{\cal O}]$ and $\nu$ are known. 

So far we have just stated the usual concept of a DIV,\cite{Ma_1976, Fisher_1983} applied to
the static scaling behavior at a quantum phase transition. Now consider the temperature 
dependence of ${\cal O}$ and define a critical exponent $\omega_T$
by ${\cal O}(r=0,T) \propto T^{\omega_T}$. Let $z$ be the dynamical exponent
as obtained from the underlying field theory by power counting. (For now we assume there is only one such $z$, we
will generalize to the case of multiple dynamical exponents below.) Strong dynamical scaling
then is expressed by a homogeneity law
\bse
\label{eqs:2.3}
\be
{\cal O}(r,T) = b^{-\omega/\nu}\,\Phi_{\cal O}(r\,b^{1/\nu}, T\,b^z)\ ,
\label{eq:2.3a}
\ee
which yields
\be
\omega_T = \omega/\nu z\ .
\label{eq:2.3b}
\ee
\ese
If an irrelevant variable is dangerous with respect to the temperature dependence of ${\cal O}$,
the relation (\ref{eq:2.3b}) no longer holds. However, as in the static case we still have a 
homogeneity law as long as weak scaling holds. We write
\bse
\label{eqs:2.4}
\be
{\cal O}(r,T) = b^{-\omega/\nu}\,\Phi_{\cal O}(r\,b^{1/\nu}, T\,b^{z_{\cal O}})\ ,
\label{eq:2.4a}
\ee
where
\be
z_{\cal O} = \omega/\nu\omega_T
\label{eq:2.4b}
\ee
\ese
with $\omega$ and $\omega_T$ the physical exponents that include the effects of the DIV.
Equations (\ref{eqs:2.4}) are valid as long as weak scaling holds. If in addition strong static
and/or dynamic scaling holds one also has the relations (\ref{eq:2.1b}) and/or (\ref{eq:2.3b})
between exponents. 

Now consider the free-energy density $f$ as a function of $r$, $T$, and the field $h$ conjugate to
the order parameter. Assuming strong scaling, we have
\be
f(r,h,T) = b^{-(d+z)}\,\Phi_f(r\,b^{1/\nu}, h\,b^{[h]}, T\,b^z)\ ,
\label{eq:2.5}
\ee
where $[h]$ is the power-counting scale dimension of $h$. We define a ``control-parameter susceptibility''
$\chi_r$ as $\chi_r = -\partial^2 f/\partial r^2$. The physical meaning of $\chi_r$ depends on the nature
of the control parameter. For instance, if the QPT is triggered by hydrostatic pressure, $\chi_r$ will be
proportional to the compressibility of the system. We further define a critical exponent $\alpha$ by 
$\chi_r(r,T=0) \propto \vert r\vert^{-\alpha}$. Note that the such-defined $\alpha$ at a QPT has
nothing to do with the specific heat; however, at a thermal transition our definition makes $\chi_r$
the specific-heat coefficient, and $\alpha$ has its usual meaning. We can now incorporate the
effect of any DIVs by writing
\be
f(r,h,T) = b^{-(2-\alpha)/\nu}\,\Phi_f(r\,b^{1/\nu}, h\,b^{\beta\delta/\nu}, T\,b^{\beta/\nu\beta_T})\ ,
\label{eq:2.6}
\ee
which is valid as long as weak scaling holds.

\subsubsection{Multiple time scales}
\label{subsubsec:II.A.2}

So far we have assumed that there is a single underlying dynamical exponent $z$ that may or
may not be modified by a DIV. An additional complication
is that often more than one dynamical exponent is present. This can happen at classical phase
transitions,\cite{DeDominicis_Peliti_1978} but it is much more common at quantum phase
transitions, chiefly because there are more soft or massless modes at $T=0$ than at $T>0$. 
For instance, at any quantum phase transition in a metallic system the coupling
of the order parameter to the conduction electrons introduces a second time scale: a ballistic
one ($z=1$) in clean systems, or a diffusive one ($z=2$) in disordered ones.\cite{Belitz_Kirkpatrick_Vojta_2005} This issue is
independent of whether or not DIVs invalidate strong scaling, so for simplicity we will assume
strong scaling in this subsection. 

For definiteness, let as assume that there are two critical time scales with dynamical exponents 
$z_1$ and $z_2 < z_1$. The generalization of the homogeneity law (\ref{eq:2.1a}) for ${\cal O}$ then reads
\be
{\cal O}(r,T) = b^{-[{\cal O}]}\,\Phi_{\cal O}(r\,b^{1/\nu}, T\,b^{z_1}, T\,b^{z_2})\ .
\label{eq:2.7}
\ee
Let us assume $[{\cal O}] > 0$ (for $[{\cal O}] < 0$ obvious modifications of the following statements hold).
Then the leading temperature dependence of ${\cal O}$ at $r=0$ will be given by $z_1$, i.e.,
$\omega_T = [{\cal O}]/z_1$. However, there is no guarantee that $\Phi(0,x,y)$ depends on $x$. 
If it does not, then $\omega_T = [{\cal O}]/z_2$, and the singular $T$-dependence of ${\cal O}$
at criticality is weaker than one would naively expect. We will see explicit examples of this in 
Sec.~\ref{sec:III}. 

For the free energy, multiple dynamical exponents result in multiple scaling parts. Let us
again consider the case of two critical time scales with dynamical exponents $z_1 > z_2$.
All of the following considerations can be trivially simplified to the case $z_1 = z_2$, and  
they are easily generalized to the case of more than two dynamical exponents. We
generalize Eq.~(\ref{eq:2.5}) to
\bea
f(r,h,T) &=& b^{-(d+z_1)}\,\Phi_f^{(1)}(r\,b^{1/\nu}, h\,b^{[h]}, T\,b^{z_1}, T\,b^{z_2})
\nonumber\\
           && + b^{-(d+z_2)}\,\Phi_f^{(2)}(r\,b^{1/\nu}, h\,b^{[h]}, T\,b^{z_2})\ .
           \nonumber\\
\label{eq:2.8}
\eea  
Note that the scaling function $\Phi_f^{(2)}$ does not depend on the argument $T\,b^{z_1}$. This is because
the entropy density, $s = \partial f/\partial T$, must vanish at least as fast as $1/\xi^d$ for $\xi\to\infty$.\cite{entropy_scaling_footnote}  
For $f$ and its derivatives with respect to $r$ and $h$ the second term obviously yields the most singular contribution. 
We thus can obtain homogeneity laws for the order parameter $m = \partial f/\partial h$, the order-parameter 
susceptibility $\chi_m = \partial^2 f/\partial h^2$, and the control-parameter susceptibility $\chi_r = \partial^2 f/\partial r^2$
from a scaling part of the free-energy density
\be
f_2(r,h,T) = b^{-(d+z_2)}\,\Phi_f^{(2)}(r\,b^{1/\nu}, h\,b^{[h]}, T\,b^{z_2})\ .
\label{eq:2.9}
\ee
For the specific-heat coefficient $\gamma = c/T = -\partial^2 f/\partial T^2$, on the other hand, the largest $z$ yields the most singular
contribution and we have
\be
\gamma(r,T) = b^{z_1 - d}\,\Phi_{\gamma}(r\,b^{1/\nu}, T\,b^{z_1}, T\,b^{z_2})\ .
\label{eq:2.10}
\ee
The leading temperature dependence at $r=0$ is in general given by the first argument of the scaling
function, i.e., $\gamma(r=0,T) \propto T^{d/z_1 - 1}$.

We emphasize again that all of the above relations assume strong scaling. See below for a discussion
of the weaker statements that remain valid if strong scaling is violated. We also stress that Eq.~(\ref{eq:2.8}),
with its two additive scaling parts, is an ansatz. In the case of the quantum ferromagnetic transition
discussed as our prime example in Sec.~\ref{sec:III} it is known explicitly that the scaling part of the free
energy has this form.
 
\subsection{Exponent relations at quantum critical points}
\label{subsec:II.B}

We now use the concepts laid out above to derive various relations between critical exponents.
           
\subsubsection{Exponent relations that rely on weak scaling}
\label{subsubsec:II.B.1}

Let the observable in question be the order parameter $m$, let $h$ be the field conjugate to
the order parameter, and let $\chi = \partial m/\partial h$ be the order-parameter susceptibility.
As long as weak scaling holds, Eqs.\ (\ref{eqs:2.4}), generalized to allow for the dependence of
$m$ on $h$, will be valid with $\omega = \beta$ and $\omega_T = \beta_T$:
\be
m(r,h,T) = b^{-\beta/\nu} \Phi_m(r\,b^{1/\nu}, h\,b^{\beta\delta/\nu}, T\,b^{z_m})\ ,
\label{eq:2.11}
\ee
where $z_m = \beta/\nu\beta_T$ is the dynamical exponent relevant for the order parameter.
In the context of Eq.~(\ref{eq:2.8}), $z_m$ is $z_2$, possibly modified by DIVs.
Differentiating with respect to $h$ we find a corresponding homogeneity law for the
order-parameter susceptibility:
\be
\chi_m(r,T) = b^{(\delta-1)\beta/\nu} \Phi_\chi(r\,b^{1/\nu}, T\,b^{z_m})\ .
\label{eq:2.12}
\ee
With the definitions of the exponents $\gamma$ and $\gamma_T$, Appendix \ref{app:A},
this implies the
\medskip
\par\noindent
{\it Widom equality:} 
\bse
\label{eqs:2.13}
\bea
\gamma &=& \beta (\delta - 1)\ ,
\label{eq:2.13a}\\
\gamma_T &=& \beta_T (\delta - 1)\ ,
\label{eq:2.13b}
\eea
\ese
which holds at a QPT for both the $r$-exponents and the $T$-exponents as a consequence
of weak scaling only. It is not affected by the presence of multiple time scales. 

Next we consider, in addition to $m$, the order-parameter susceptibility $\chi_m$ and the
control-parameter susceptibility $\chi_r$, which all are obtained from the scaling part $f_2$
of the free-energy density, Eq.~(\ref{eq:2.9}). Incorporating the effects
of DIVs, if any, as in Sec.~\ref{subsubsec:II.A.1}, we can write
\bse
\label{eqs:2.14}
\bea
m(r,h,T) &=& b^{(\beta\delta + \alpha - 2)/\nu}\,\Phi_m(r\,b^{1/\nu}, h\,b^{\beta\delta/\nu}, T\,b^{z_m})\ ,
\nonumber\\
\label{eq:2.14a}\\
\chi_m(r,T) &=& b^{(2\beta\delta + \alpha - 2)/\nu}\,\Phi_{\chi}(r\,b^{1/\nu}, T\,b^{z_m})\ ,
\label{eq:2.14b}\\
\chi_r(r,T) &=& b^{\alpha/\nu}\,\Phi_r(r\,b^{1/\nu}, T\,b^{z_m})\ .
\label{eq:2.14c}
\eea
\ese
From the definitions of the exponents $\beta$, $\gamma$, $\beta_T$, $\gamma_T$, and $\alpha_T$, 
and using Eq.~(\ref{eq:2.13a}) we then obtain the
\medskip
\par\noindent
{\it Essam-Fisher equality:} 
\bse
\label{eqs:2.15}
\be
\alpha + 2\beta + \gamma = 2\ ,
\label{eq:2.15a}
\ee
for the $r$-exponents, and its analog
\be
\alpha_T + 2\beta_T + \gamma_T = 2/\nu z_m
\label{eq:2.15b}
\ee
\ese
for the $T$-exponents. Equation~(\ref{eq:2.15a}) depends on
weak scaling only. Note that $\alpha$ is not the specific-heat exponent,
but rather the control-parameter-susceptibility exponent defined in Sec.~\ref{subsubsec:II.A.1} and
Appendix~\ref{app:A}. Equation~(\ref{eq:2.15a}) is the natural extension of the classical Essam-Fisher
equality to quantum phase transitions, even though it does not involve a specific-heat exponent. 
The simple relation between Eq.~(\ref{eq:2.15a}) and (\ref{eq:2.15b}), which is consistent with
Eq.~(\ref{eq:2.4b}) with $z_{\cal O} = z_m$, is due to the fact that the dominant dynamical exponent
is the same for all three observables $m$, $\chi_m$, and $\chi_r$. The latter can be guaranteed
only in the presence of strong scaling, see the remarks after Eq.~(\ref{eq:2.8}). However, while
strong scaling is sufficient for Eq.~(\ref{eq:2.15b}), it is not necessary.

Equation~(\ref{eq:2.14a}) also implies $-\beta = \beta\delta + \alpha - 2$. Together with the Essam-Fisher
equality this allows us to express $\delta$ in terms of $\alpha$ and $\gamma$:
\be
\delta = \frac{2 - \alpha + \gamma}{2 - \alpha - \gamma}\ .
\label{eq:2.16}
\ee
This extends to QPTs another relation that is well known for thermal phase transitions.\cite{Stanley_1971}
Note that Eq.~(\ref{eq:2.16}) is not independent; it follows from a combination of the Widom and 
Essam-Fisher equalities. We also stress again that in the current context $\alpha$ is not the 
specific-heat coefficient.

Now consider the order-parameter two-point correlation function 
$G({\bm x}) = \int_0^{\infty} d\tau\,\langle m({\bm x},\tau)\,m({\bm 0},0)\rangle$, with $\tau$ the 
imaginary-time variable. The spatial integral of $G({\bm x})$ yields the
order-parameter susceptibility, $\chi_m = \int d{\bm x}\ G({\bm x})$. It has the form
\be
G({\bm x}) = \frac{e^{-\vert{\bm x}\vert/\xi}}{\vert{\bm x}\vert^{d-2+\eta}}\ ,
\label{eq:2.17}
\ee
which defines both the correlation length $\xi$ and the critical exponent $\eta$. One thus
has
\be
\chi_m \propto \xi^{2-\eta}\ .
\label{eq:2.18}
\ee
An equivalent argument is to generalize the homogeneity equation for the order-parameter
susceptibility to include the wave-number dependence:
\be
\chi_m(r,T;k) = b^{\gamma/\nu}\,\Phi_{\chi}(r\,b^{1/\nu}, T\,b^{z_m}; k\,b)\ .
\label{eq:2.19}
\ee
With the definitions of the exponents $\nu$ and $\nu_T$, Appendix~\ref{app:A}, we obtain
from either Eq.~(\ref{eq:2.18}) or (\ref{eq:2.19}) the
\medskip
\par\noindent
{\it Fisher equality:}
\bse
\label{eqs:2.20}\bea
\gamma &=& (2 - \eta)\nu\ ,
\label{eq:2.20a}\\
\gamma_T &=& (2 - \eta)\nu_T
\label{eq:2.20b}
\eea
\ese
for both the $r$-exponents and the $T$-exponents. It depends on weak scaling only. 

In summary, we have found three independent weak-scaling exponent relations for
the $r$-exponents, viz. the Widom, Fisher, and Essam-Fisher equalities (\ref{eq:2.13a}),
(\ref{eq:2.15a}), and (\ref{eq:2.20a}), and three corresponding relations for the 
$T$-exponents, Eqs.~(\ref{eq:2.13b}), (\ref{eq:2.15b}), and (\ref{eq:2.20b}).

\subsubsection{A rigorous inequality}
\label{subsubsec:II.B.2}

Exponent relations that involve the specific-heat exponent $\bar\alpha$ (see the Introduction and
Appendix~\ref{app:A} for a definition and discussion of $\bar\alpha$) fall into a different class,
since the critical behavior of the specific-heat coefficient is in general governed by a scaling part
of the free energy that is different from the one that determines the exponents discussed so far,
see Eq.~(\ref{eq:2.10}) and the related discussion. We start with the thermodynamic identity 
that underlies the classical Rushbrooke inequality\cite{Rushbrooke_1963b, Stanley_1971}
\be
1 - \gamma_m/\gamma_h = \left[(\partial m/\partial T)_h^2\right]/\chi_T\,\gamma_h\ .
\label{eq:2.21}
\ee
Here $\gamma_m$ and $\gamma_h$ denote the specific-heat coefficient at fixed order parameter
and fixed conjugate field, respectively, and $\chi_T$ is the isothermal OP susceptibility. Note
that, in order to apply this relation to QPTs, it is crucial to formulate it in terms of the specific-heat
coefficients; the usual formulation in terms of the specific heats leads to a factor of $T$ on the
right-hand side that makes the $T\to 0$ limit ill defined.~\cite{gamma_footnote} With the $T$-exponents
as defined in Appendix~\ref{app:A} this yields an exponent inequality at a QPT in the form
\bse
\label{eqs:2.22}
\be
{\bar\alpha}_T + 2\beta_T + \gamma_T \geq 2\ .
\label{eq:2.22a}
\ee
This is rigorous, since it depends only on thermodynamic stability arguments. If in addition we use
weak scaling of the order parameter, Eq.\ (\ref{eq:2.11}), we find a corresponding inequality for the
$r$-exponents,
\be
{\bar\alpha} + 2\beta + \gamma \geq 2\beta/\beta_T\ .
\label{eq:2.22b}
\ee
\ese
This follows since weak scaling of the OP implies $(\partial m/\partial T)_{T=0,h=0} \propto (-r)^{\beta(1-1/\beta_T)}$.
Note that this is different from the classical Rushbrooke inequality, Eq.\ (\ref{eq:B.5}). As for the latter,
Eqs.\ (\ref{eqs:2.22}) hold as equalities if $\gamma_m/\gamma_h \to 1$ for $T\to 0$ at $r=0$, and
for $r\to 0$ at $T=0$, respectively, and as inequalities otherwise. In the classical limit, both of the 
Eqs.~(\ref{eqs:2.22}) turn into the the classical Rushbrooke inequality: The $T$-exponents coincide with the
$r$-exponents, and $\bar\alpha$ coincides with the classical specific-heat exponent $\alpha$, see the
discussion in Appendix~\ref{app:A}.

\subsubsection{Hyperscaling relations}
\label{subsubsec:II.B.3}

The exponent relations we derived and discussed so far depended at most on weak scaling (with the
exception of Eq.~(\ref{eq:2.15b}), which relies to some extent on a strong-scaling assumption). If strong
scaling is valid we can derive additional constraints on the exponents. For this purpose, we return 
to Eq.~(\ref{eq:2.8}). As we discussed in this context, $z_2 = z_m$ is the dynamical exponent for
the order parameter, the order-parameter susceptibility, and the control-parameter susceptibility,
while $z_1 = z_c \geq z_m$ is the dynamical exponent for the specific-heat coefficient. In general
$z_c$ could be $z_1$ modified by DIVs, but the strong-scaling assumption means $z_c = z_1$. Strong scaling
thus implies the quantum versions of the usual hyperscaling relation  (\ref{eq:B.1d}),
\bse
\label{eqs:2.23}
\bea
\alpha &=& 2 - \nu(d + z_m)\ ,
\label{eq:2.23a}\\
\alpha_T &=& 2/\nu z_m - d/z_m - 1\ .
\label{eq:2.23b}
\eea
\ese
The latter in agreement with Eq.~(\ref{eq:2.3b}), and the former (for the case where there is only
one dynamical critical exponent) with Ref.~\onlinecite{Continentino_Japiassu_Troper_1989}. We will refer to these as the $\alpha$-hyperscaling relations.
Analogously, we find from the first term in Eq.~(\ref{eq:2.8}) hyperscaling relations for the specific-heat exponents
\bse
\label{eqs:2.24}
\bea
{\bar\alpha} &=& \nu (z_c - d)\ ,
\label{eq:2.24a}\\
{\bar\alpha}_T &=& 1 - d/z_c\ ,
\label{eq:2.24b}
\eea
\ese
which we will refer to as the ${\bar\alpha}$-hyperscaling relations. 

Equations~(\ref{eqs:2.23}) and (\ref{eqs:2.24}) are two independent strong-scaling relations. In conjunction
with the weak-scaling relations from Sec.~\ref{subsubsec:II.B.1} we can use them to derive additional
relations that are not independent. For instance, Eqs.~(\ref{eqs:2.23}) and (\ref{eqs:2.24}) imply a
relation between the control-parameter
exponent $\alpha$ and the specific-heat exponent ${\bar\alpha}$,
\bse
\label{eqs:2.25}
\bea
{\bar\alpha} &=& \alpha - 2 + \nu(z_m + z_c)\ ,
\label{eq:2.25a}\\
{\bar\alpha}_T &=& \alpha_T - 2/\nu z_m + 2 + d(1/z_m - 1/z_c)\ .\quad
\label{eq:2.25b}
\eea
\ese
Combining these with the Essam-Fisher relations (\ref{eqs:2.15}) we find
\bse
\label{eqs:2.26}
\bea
{\bar\alpha} + 2\beta + \gamma &=& \nu(z_m + z_c)\ ,
\label{eq:2.26a}\\
{\bar\alpha}_T + 2\beta_T + \gamma_T &=& 2 + d (1/z_m - 1/z_c)\ ,
\nonumber\\
\label{eq:2.26b}
\eea
\ese
Both of these equalities are consistent with Eqs.~(\ref{eqs:2.22}),
since $z_c \geq z_m$ always. Note that they are physically different from the Essam-Fisher
equalities (\ref{eqs:2.15}), which depend on weak scaling only. 

We can use Eq.~(\ref{eq:2.25a}) together with Eq.~(\ref{eq:2.16}) to express $\delta$ in terms of $\bar\alpha$, $\gamma$,
and the dynamical exponents,
\be
\delta = \frac{\nu(z_m + z_c) - {\bar\alpha} + \gamma}{\nu(z_m + z_c) - {\bar\alpha} - \gamma}\ .
\label{eq:2.27}
\ee
We note that all of the above relations rely on strong scaling, even if they do not explicitly involve the
dimensionality.

Finally, the electrical conductivity is dimensionally an inverse length to the power $(d-2)$. A strong-scaling hypothesis
thus implies that the scaling part $\Delta\sigma$ of the conductivity obeys a homogeneity law
\be
\Delta\sigma(r,T) = b^{-(d-2)}\,F_{\sigma}(r\,b^{1/\nu}, T\,b^{z_c})\ .
\label{eq:2.28}
\ee
For the exponents $s$ and $s_T$ defined in Eq.~(\ref{eq:A.6}) this implies the
\medskip\par\noindent
{\it Wegner equality:}
\bse
\label{eqs:2.29}
\bea
s &=& \nu(d-2)\ ,
\label{eq:2.29a}\\
s_T &=& (d-2)/z_c\ .
\label{eq:2.29b}
\eea
\ese
These relations play an important role in the theory of electron localization,\cite{Wegner_1976a, Abrahams_et_al_1979}
but are much more generally applicable. They do, however, depend on strong scaling and are not valid if DIVs affect
the conductivity. Note that in general, with DIVs taken into account, $s$ and $s_T$ can be either positive or negative
in any dimension, with negative values applying to either the clean limit or regimes where the residual resistivity is
small compared to the temperature-dependent part.\cite{conductivity_scaling_footnote}

\subsection{Scaling at quantum first-order transitions}
\label{subsec:II.C}

Fisher and Berker have shown how a classical first-order transition can be understood within a standard scaling and RG 
framework.\cite{Fisher_Berker_1982} In a RG context, at a first-order transition all gradient-free operators in a 
Landau-Ginzburg-Wilson (LGW) theory are relevant and 
have the largest scale dimension that is thermodynamically allowed, viz., $d$. This just follows from the fact that
the OP itself is dimensionless at a first-order transition, so the operator dimensions must make up for the scale dimension
of the spatial integration measure.\cite{Landau_expansion_footnote} In particular, $\nu = 1/d$. This is reconciled with the usual notion
of a finite correlation length at a first-order transition by means of finite-size scaling considerations. A dimensionless order
parameter requires $[h] = d$, which is equivalent to $\beta = 0$, and the remaining classical exponent values follow
readily: $\eta = 2-d$, $\gamma=1$, $\delta = \infty$, and $\alpha = 1$. All scaling relations, Appendix \ref{app:B},
are fulfilled, including the hyperscaling relation. (Note that DIVs are not an issue in this case, since the suspect operators are 
relevant.) 

The reasoning of Fisher and Berker can readily be extended to QPTs.\cite{finite_size_scaling_footnote}
For the case of making the logic clear, we
first do so for a QPT with only one dynamical exponent, and then generalize to the case of two dynamical exponents. Consider again the homogeneity law for 
the free energy, Eq.\ (\ref{eq:2.5}). In order for the order parameter $m = \partial f/\partial h$ to be dimensionless we must have 
\be
[h] = d+z\ . 
\label{eq:2.30}
\ee
A dimensionless OP in turn implies
\bse
\label{eqs:2.31}
\be
1/\nu = d + z\ ,
\label{eq:2.31a}
\ee
which generalizes the classical $1/\nu = d$.\cite{Continentino_Ferreira_2004} It further implies
\be
\eta = 2 - d - z\quad,\quad \delta = \infty\quad,\quad \beta = 0\quad,\quad \beta_T = 0\ .
\label{eq:2.31b}
\ee
Considering the OP susceptibility $\chi_m = \partial^2 f/\partial h^2$, and using Eqs.~(\ref{eq:2.30}) and (\ref{eq:2.31a}), we have
\be
\gamma = 1\quad,\quad \gamma_T = (d + z)/z\ .
\label{eq:2.31c}
\ee
Similarly, we find from the homogeneity law for the control-parameter susceptibility $\chi_r = \partial^2 f/\partial r^2$ the exponents
\be
\alpha = 1 \quad,\quad \alpha_T = (d+z)/z\ .
\label{eq:2.31d}
\ee
The value $\alpha = 1$ implies that the first derivative of the free energy with respect to $r$ is discontinuous, and the second
derivative contains a delta-function. This is analogous to the entropy being discontinuous, and the existence of a latent heat,
at a thermal phase transition. In the quantum case the physical interpretation of $\alpha = 1$ depends on the nature of the
control parameter; see the discussion in Sec.~\ref{subsubsec:IV.B.1}. From the homogeneity law for the 
specific-heat coefficient $\gamma = \partial^2 f/\partial T^2$ we find
\be
{\bar\alpha} = \frac{z - d}{z + d}\quad,\quad \bar\alpha_T = \frac{z - d}{z}\ .
\label{eq:2.31e}
\ee
Finally, Eq.\ (\ref{eq:2.2}) holds in particular for the correlation-length exponent, which yields
\be
\nu_T = 1/z\ .
\label{eq:2.31f}
\ee
\ese
We see that all scaling relations discussed in Sec.\ \ref{subsec:II.A} hold, in analogy to the classical case.

In the above discussion we have assumed what is called a {\em block}
geometry in the finite-size scaling theory of classical first order
phase transitions. Physically it is realized in a finite system whose linear sizes in all
dimensions, including the imaginary-time or inverse temperature one, are comparable. 
Another choice is a {\em cylinder} geometry in which the linear size in one dimension is
large compared to the others. In the quantum case this is of special interest for a system
with fixed finite volume in the zero-temperature limit, where the size in the imaginary-time direction
becomes infinitely large. In this case arguments identical to those used in classical finite-size
scaling theory \cite{Privman_Fisher_1983} lead to a time scale that scales as
\be
\tau \sim e^{\,\sigma L^d}\ ,
\label{eq:2.32}
\ee
with $\sigma$ a positive constant. In the classical case the analogous
result has led to a number of remarkable conclusions that have been
confirmed experimentally.\cite{Abraham_Maciolek_Vasilyev_2014} The
physical meaning of $\tau$ in the quantum case is the time it takes for
a droplet of volume $L^d$ containing a certain state to transform into a different state
via a tunneling process. 

Now we turn to the case of two different dynamical exponents for the OP and the specific heat, which we again
denote by $z_m$ and $z_c$, respectively. The imaginary-time integral in the mass term and the Zeeman term
in a LGW functional will then have a scale dimension of $-z_m$ (taking the scale dimension of a length to
be $-1$). Making the OP dimensionless thus requires
\bse
\label{eqs:2.33}
\be
[h] = d + z_m\ ,
\label{eq:2.33a}
\ee
and 
\be
1/\nu = d + z_m\ .
\label{eq:2.33b}
\ee
The OP is now dimensionless by construction, which implies
\be
\eta = 2 - d - z_m\quad,\quad \delta = \infty\quad,\quad \beta = 0\quad,\quad\beta_T = 0\ .
\label{eq:2.33c}
\ee
From the OP susceptibility, $\chi_m = \partial m/\partial h$, we obtain
\be
\gamma = 1\quad,\quad\gamma_T = (d + z_m)/z_m\ ,
\label{eq:2.33d}
\ee
and from the control-parameter susceptibility $\chi_r = \partial^2 f/\partial r^2$
\be
\alpha = 1\quad,\quad \alpha_T = (d + z_m)/z_m\ .
\label{eq:3.32e}
\ee
Since the correlation length is entirely a property of the OP-OP correlation function, its temperature
dependence is also governed by $z_m$ and Eq.\ (\ref{eq:2.4b}) holds for $\omega = \nu$ with $z = z_m$:
\be
\nu_T = 1/z_m\ .
\label{eq:2.33f}
\ee
The specific-heat coefficient is governed by the dynamical exponent $z_c$, and we thus have
\be
{\bar\alpha} = \nu(z_c-d) = \frac{z_c-d}{z_m+d}\quad,\quad{\bar\alpha_T} = 1 - d/z_c\ .
\label{eq:2.33g}
\ee
\ese
All of the scaling relations discussed in Sec.\ \ref{subsec:II.A} still hold. 

We will see explicit examples for the general relations developed here in Sec.\ \ref{sec:III}, and 
we will further discuss the underlying concepts and their consequences in Sec.\ \ref{sec:IV}.

\section{Application to quantum ferromagnets}
\label{sec:III}

The scaling theory developed in Sec.~\ref{sec:II} is in general quite complex
because of the occurrence of multiple time scales at many quantum
phase transitions. As an explicit illustration we now apply the
theory to the quantum ferromagnetic phase transitions
in metallic systems. For a review of the ferromagnetic quantum phase transition problem,
see Ref.~\onlinecite{Brando_et_al_2015}.

\subsection {Clean systems}
\label{subsec:III.A}

The ferromagnetic transition in metals was one of the earliest quantum phase transitions
considered, and it was used as an example by Hertz in his seminal paper on a renormalization-group
approach to quantum phase transitions.\cite{Hertz_1976} Hertz considered a ferromagnetic order parameter ${\bm m}$ 
coupled to conduction electrons. The latter lead to Landau damping of the magnetic fluctuations, and
on general grounds, using symmetry arguments as well as the known structure of the Fermi-liquid dynamics,
one can write down an action
\bea
{\cal A}_{\text{Hertz}} &=& -\sum_{{\bm k},n} \left[r + a{\bm k}^2 + c\vert\Omega_n\vert/\vert{\bm k}\vert\right]
\nonumber\\
&&\times {\bm m}({\bm k},\Omega_n) \cdot {\bm m}({\bm k},-\Omega_n)
                                      + O({\bm m}^4)\ ,
\label{eq:3.1}
\eea
where we have written only the term quadratic in the order parameter explicitly. Here $\Omega_n$ is a bosonic
Matsubara frequency, and $r$, $a$, and $c$ are parameters of the LGW functional. The term $\vert\Omega_n\vert/\vert{\bm k}\vert$
reflects the dynamics of the conduction electrons. This is the action considered by Hertz, who explicitly derived it
from a specific model. From this he concluded that the dynamical critical exponent
is $z=3$ (since $\Omega \sim \vert{\bm k}\vert^3$ for $r=0$), and that the upper critical dimension is                                       
$d_{\text{c}}^+ = 4 - z = 1$. This in turn led to the conclusion that the quantum phase transition is second order
and described by a simple Gaussian fixed point, and that the static critical behavior is mean-field-like
for all $d\geq 1$ with logarithmic corrections to scaling for $d=1$.\cite{ucd_footnote}  The behavior at finite temperature was considered
in detail by Millis,\cite{Millis_1993} and the results of the RG treatment for $d=3$ confirmed results obtained earlier by Moriya and
co-workers by means of what is often called self-consistent spin-fluctuation theory.\cite{Moriya_1985} This combined body of work
is often referred to as Hertz-Millis-Moriya theory.

It was later shown that the above conclusions do not hold due to properties of the conduction electrons that are
not captured in Hertz's action. A careful analysis of the soft fermionic modes coupling to the magnetization shows
that, in three-dimensional systems, the leading wave-vector dependence in the Gaussian action is a term proportional to 
${\bm k}^2\ln(1/\vert{\bm k}\vert)$ with a {\em negative} prefactor,\cite{Belitz_Kirkpatrick_Vojta_1997} and such a term is indeed generated under
renormalization if one starts with Hertz's action. An equivalent statement is that there is a $m^4\ln(1/m)$ term in a
generalized Landau theory for the ferromagnetic quantum phase transition.\cite{Brando_et_al_2015}
As a result, Hertz's fixed point is not stable, and the ferromagnetic quantum phase transition
in clean metals is generically first order.\cite{Belitz_Kirkpatrick_Vojta_1999, Kirkpatrick_Belitz_2012b} However, depending on quantitative
details involving the strength of the electron correlations and the inevitable weak 
disorder, there can be a sizable regime where the critical behavior associated with Hertz's fixed point is
observable, even though asymptotically close to the transition it crosses over to a first-order transition.\cite{Brando_et_al_2015}
In this subsection we therefore discuss, and include in Table~\ref{table:1}, the critical exponents at
Hertz's fixed point, together with the exponents associated with the ultimate first-order transition, and
the critical behavior at the termination point of the tricritical wings that result from the first-order transition.\cite{Belitz_Kirkpatrick_Rollbuehler_2005}

\begin{figure*}[t]
\begin{center}
\hspace*{-25pt}
\includegraphics[width=2.0\columnwidth,angle=0]{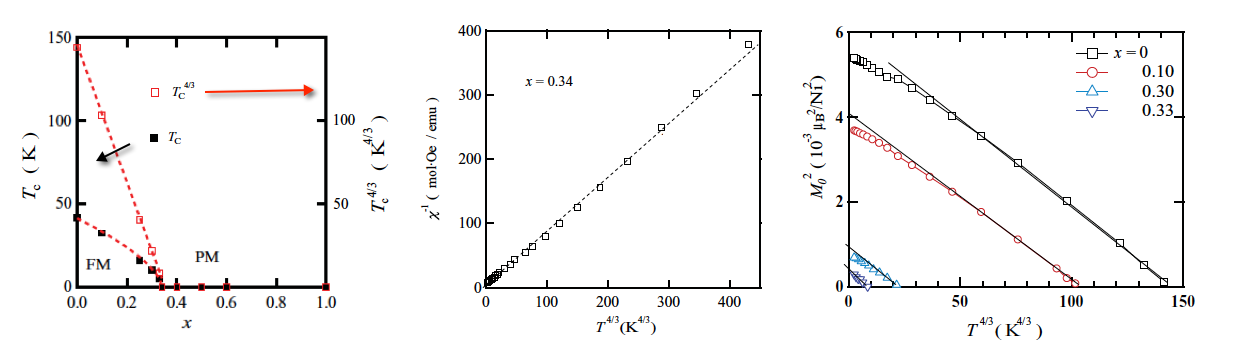}
\end{center}
\vskip -20pt
\caption{Hertz-type scaling behavior as observed in Ni$_3$Al$_{1-x}$Ga$_2$. From left to right: $T_{\text{c}}$ vs $x$ phase diagram, inverse magnetic susceptibility 
              vs. $T^{4/3}$ for the critical sample, magnetization squared vs. $T^{4/3}$ for various concentrations. Figure adapted from Ref.~\onlinecite{Yang_et_al_2011}.}
\label{fig:Ni3AlGa}
\end{figure*}

\subsubsection{Hertz's fixed point}
\label{subsubsec:III.A.1}

Table~\ref{table:1} lists the critical exponents associated with Hertz's fixed point.
The values for the static exponents are straightforward generalizations of the results obtained in Refs.~\onlinecite{Hertz_1976, Millis_1993}.
The value for the resistivity exponent $s_T$ is Mathon's result\cite{Mathon_1968} generalized to $d$ dimensions, and $z_c = 3$ is Hertz's value for the
dynamical critical exponent.\cite{Hertz_1976} There was no notion of a separate dynamical exponent $z_m$ in the original work.
However, this concept naturally arises if we take the point of view expressed in Eqs.~(\ref{eqs:2.4}) and incorporate the effects of the
DIVs in the homogeneity laws. Equation~(\ref{eq:2.4b}) applied to the order parameter then implies $z_m = \beta/\nu\beta_T$, which
yields $z_m = 1/\beta_T$ as listed in Table~\ref{table:1}. Alternatively, one can work with only one dynamical exponent, $z=3$, and consider 
the DIV explicitly, as was done by Millis,\cite{Millis_1993} see also Ref.~\onlinecite{Brando_et_al_2015}. This reasoning effectively leads to
$z_m = z/(1 + \nu(d-1))$, which is the same result as above. At the upper critical dimension
$d=1$ the effects of the DIVs disappear, and there is only one dynamical exponent $z_m = z_c = 3$. Note that there is only one fundamental
time scale in Hertz's theory, and the appearance of a second dynamical exponent is solely due to the fact that the fundamental $z$ is
modified by a DIV in some contexts, but not in others. This is very different from the theories discussed in Secs.~\ref{subsubsec:III.A.2}, \ref{subsubsec:III.A.3}, 
and \ref{subsubsec:III.B.2}, which intrinsically contain two time scales of different physical origin. This is an indication
that important physics related to the conduction electrons is missing in Hertz's theory; see Sec.~\ref{subsubsec:IV.B.2} for additional comments
on this issue.

Since for $d>1$ the system is above its upper critical dimension, DIVs in general
invalidate any scaling relations that rely on strong scaling as defined in Sec.~\ref{subsec:II.A},
while those that rely on weak scaling only will remain valid. Indeed, considering the values in the table, 
we see that the Widom equalities (\ref{eqs:2.13}), the Essam-Fisher equalities (\ref{eqs:2.15}), the Fisher equalities (\ref{eqs:2.20}), and the
Rushbrooke inequalities (\ref{eqs:2.22}) are satisfied for all $d\geq 1$. The hyperscaling relations (\ref{eqs:2.23}), on the other hand, do not hold. 
The hyperscaling relations (\ref{eqs:2.24}) are a more complicated case. The values of ${\bar\alpha}$ and ${\bar\alpha}_T$ listed
in Table~\ref{table:1} reflect the leading fluctuation contribution to the specific-heat coefficient, which do obey strong scaling. For
$1\leq d <3$ they dominate the constant mean-field contribution which violates strong scaling, and as a result Eqs.~(\ref{eqs:2.24})
hold even though the system is above its upper critical dimension. For $d>3$ they are subleading compared to the
mean-field contribution, the specific-heat exponents lock into their mean-field values ${\bar\alpha} = {\bar\alpha}_T = 0$, and hyperscaling breaks down. 

The contribution to the resistivity $\rho$ due to scattering of electrons by critical fluctuations was calculated by Mathon,\cite{Mathon_1968} who found
$\Delta\rho(r=0,T) \propto T^{5/3}$ in $d=3$. A simple generalization to general $d$ yields $\Delta\rho(r=0,T) \propto T^{(d+2)/3}$, or $s_T = -(d+2)/3$ for the
conductivity exponent $s_T$ defined in Appendix~\ref{app:A}. From a scaling point of view, this result can be made plausible as follows. Consider
the strong-scaling homogeneity law (\ref{eq:2.28}) for the conductivity contribution $\Delta\sigma$. In a clean system, the backscattering factor in the Boltzmann equation
provides an additional factor of the hydrodynamic momentum squared, which is not captured by power counting. This leads to a scale dimension
of $\Delta\sigma$ that is effectively equal to $d-4$ rather than $d-2$. In addition, the DIV $u$ with scale dimension $[u] = -(d-1)$ affects $\Delta\sigma$, and
from Fermi's golden rule it is plausible that $\Delta\sigma \propto 1/u^2$. Combining these arguments, we have
\bea
\Delta\sigma(r,T) &=& b^{-(d-4)} b^{2(d-1)}\,F_{\sigma}(r\,b^{1/\nu}, T\,b^{z_c})
\nonumber\\
                           &=& b^{(d+2)}\,F_{\sigma}(r\,b^{1/\nu}, T\,b^{z_c})\ ,
\label{eq:3.2}
\eea
which yields $s_T = -(d+2)/z_c$ as quoted above and listed in Table~\ref{table:1}. Note that the Wegner equality (\ref{eq:2.29b}) is violated for
two reasons: First, the conductivity, or the underlying relaxation rate, does not have its power-counting scale dimension due to the backscattering
factor in the Boltzmann equation. Second, the DIV further modifies the scale dimension and actually changes its sign for all dimensions. Finally, note
that the exponent $s$ as defined in Eq.~(\ref{eq:A.6}) does not exist since $\sigma(T=0) = \infty$. Rather, for $T\to 0$ at fixed $r\neq 0$ the 
conductivity will cross over to the Fermi-liquid result, $\Delta\sigma(T) \propto T^{-2}$, and Eq.~(\ref{eq:3.2}) yields the $r$-dependence of the
prefactor:
\be
\Delta\sigma(r\neq 0,T\to 0) \propto \vert r\vert^{(4-d)/2}\,T^{-2}\ .
\label{eq:3.2'}
\ee
                                             
As an illustration, we discuss an experiment on the ferromagnet Ni$_3$Al$_{1-x}$Ga$_x$, which displays a QPT at $x = x_c \approx 0.34$.\cite{Yang_et_al_2011}
This system is only moderately disordered, and critical behavior associated with the clean Hertz fixed point is expected to be observable in a sizable
transient regime.\cite{Brando_et_al_2015} There are three experimental observations, from which four exponents can be deduced: (1)~The critical 
temperature scales as $T_{\text{c}} \sim r^{3/4}$, see the first panel in Fig.~\ref{fig:Ni3AlGa}. This implies $\nu z_m = \nu z/(1 + 2\nu) = 3/4$, as can be seen, 
for instance, from Eq.~(\ref{eq:2.12}): The critical temperature is determined by $\chi_m$ diverging, which must happen for a particular value of 
the argument $x$ in $\Phi_{\chi}(1,x)$. This in turn implies $T_{\text{c}}/r^{\nu z_m} = \text{const.}$ (2)~At the critical concentration, the magnetic
susceptibility scales as $\chi_m \sim T^{-4/3}$, see the second panel in Fig.~\ref{fig:Ni3AlGa}. This implies $\gamma_T = 4/3$. Together with the first observation, 
it also implies $\gamma = 1$, since $r \sim T^{4/3} \sim 1/\chi$. (3)~The magnetization vanishes as $m^2 \propto T_{\text{c}}^{4/3} - T^{4/3}$, see the third
panel in Fig.~\ref{fig:Ni3AlGa}. In addition to confirming the product $\nu z_m = 3/4$ this yields $\beta = 1/2$, see the discussion after Eq.~(\ref{eq:4.2}). 
All of these results are in agreement with the theoretical results summarized in Table~\ref{table:1}, 
specialized to $d=3$. The conductivity exponent $s_T = -5/3$ has also been observed in various materials, see, e.g., Ref.~\onlinecite{Moriya_1985}.

\begin{table*}[p!]
\caption{Static, dynamic, and transport critical exponents as defined in the text, and validity of exponent relations, at various fixed points for the quantum 
              phase transition in metallic 
              ferromagnets. The first two columns of values represent true asymptotic exponents (QWCP = quantum wing-critical point). The third and fourth columns of values refer
               to critical behavior that is asymptotically 
              not represented by pure power laws, and the fifth and sixth columns of values represent power-law behavior that is observable only in a transient non-asymptotic 
              regime. Strong-scaling relations for the $T$-exponents are not shown,
              they hold if an only if the corresponding relations for the $r$-exponents hold. See the text for additional explanation. 
              N/A = not applicable. }
\vskip 10pt
\begin{ruledtabular}
\begin{tabular}{| c | cc | c | cc | cc | cc}
 & \multicolumn{2}{c}{\mbox{ }}  & &  \multicolumn{6}{c}{\hskip -10pt Ferromagnetic Fixed Point}\\[2pt]
\cline{5-10}\\[-7pt]
& \multicolumn{2}{c}{ } &  & & \multicolumn{2}{c}{\hskip 0pt True Critical Fixed Points} & &  \multicolumn{2}{c} {Unstable Fixed Points} \\[2pt]
 \cline{5-10}
& \multicolumn{2}{c}{ } &&&&&&&\\[-7pt]
& \multicolumn{2}{c}{ } && clean 
      & clean 
         & dirty$\,^d$ 
            & dirty (pre-
               & clean
                  & dirty \\ 
& \multicolumn{2}{c}{ } &
   & (1$^\text{st}$ order)$^a$
      & (QWCP)$^{b,c}$
         & 
            &  asymptotic)$\,^{e}$   
               &  (Hertz)$\,^{c}$                     
                  & (Hertz)$\,^{f}$ \\ 
&&&&&&&&&\\[-7pt]                    
\cline{1-10}
&&&&&&&&&\\[-2pt]   
\mbox{\hskip 15pt}&\mbox{\hskip 10pt} & \mbox{\hskip 10pt} & $\nu$ 
   & $\frac{1}{d+1}$
      & $\frac{1}{2}$
         & $\frac{1}{d-2}$
            & $\frac{1}{d-2+\lambda}$
               & $\frac{1}{2}$
                  & $\frac{1}{2}$\\
 & & &  & & & & & \\[-5pt]
&&&$\nu_T$
   & $1$
      & $\frac{d+1}{6}$
         & $\frac{1}{2}$
            & $\frac{1}{2 + \lambda}$
               & $\frac{d+1}{6}$
                  & $\frac{d+2}{8}$\\  
&&&&&&&&&\\[-5pt]                                                   
&&&$\beta$
   & 0
      & $\frac{1}{2}$
         & $\frac{2}{d-2}$
            & $\frac{2}{d-2+\lambda}$
               & $\frac{1}{2}$
                  & $\frac{1}{2}$ \\
&&&&&&&&\\[-5pt]                  
&&& $\beta_T$
   & 0
      & $\frac{d+1}{6}$
         & 1
            & $\frac{2}{2+\lambda}$ 
               & $\frac{d+1}{6}$
                  & $\frac{d+2}{8}$ \\ 
&&&&&&&&&\\[-5pt]  
&&&$\delta$
   & $\infty$
      & $3$
         & $\frac{d}{2}$
            & $\frac{d+\lambda}{2}$
               & 3
                  & 3 \\
&&&&&&&&&\\[-5pt] 
&&\begin{rotate}{90} \hskip -15pt static \end{rotate}
&$\gamma$
   & $1$
      & $1$
         & $1$
            & $1$
               & 1
                  & 1 \\
&&&&&&&&&\\[-5pt] 
&& & $\gamma_T$
   & $d+1$
      & $\frac{d+1}{3}$
         & $\frac{d-2}{2}$
            & $\frac{d-2+\lambda}{2+\lambda}$
               & $\frac{d+1}{3}$
                  & $\frac{d+2}{4}$ \\ 
&&&&&&&&&\\[-5pt]  
&&&$\eta$
   & $1-d$
      & $0$
         & $4-d$
            & $4-d-\lambda$
               & 0
                  & 0 \\
&&&&&&&&&\\[-5pt]    
&&&${\alpha}$
   & $1$
      & $0$
         &  $\frac{d-6}{d-2}$
            & $\frac{d-6+\lambda}{d-2+\lambda}$
               & $0$
                  & $0$ \\
&&&&&&&&&\\[-5pt]      
\begin{rotate}{90}  Exponents  \end{rotate} 
& &&${\alpha}_T$
   & $d+1$
      & $0$
         & $\frac{d-6}{2}$
            & $\frac{d-6+\lambda}{2+\lambda}$
               & $0$
                  & $0$ \\ 
&&&&&&&&&\\[-5pt] 

&&&${\bar\alpha}$
   & $0$
      &  $\frac{3-d}{2}$ or $0$$\,^{g}$
         & $0$
            & $\frac{\lambda}{d-2+\lambda}$
               & $\frac{3-d}{2}$ or $0$$\,^{g}$
                  & $\frac{4-d}{2}$ or $0$$\,^{h}$ \\
&&&&&&&&&\\[-5pt]                  
& &&${\bar\alpha}_T$
   & $0$
      & $\frac{3-d}{3}$ or $0$$\,^{g}$
         & $0$
            & $\frac{\lambda}{d+\lambda}$
               & $\frac{3-d}{3}$ or $0$$\,^{g}$
                  & $\frac{4-d}{4}$ or $0$$\,^{h}$ \\ 
&&&&&&&&&\\[-5pt] 
\cline{2-10} 
&&&&&&&&&\\[-5pt]   
& 
&&$z_m$
   & $1$
      & $\frac{6}{d+1}$
         & $2$
            & $2 + \lambda$
               & $\frac{6}{d+1}$
                  &  $\frac{8}{d+2}$\\
&&&&&&&&&\\[-5pt]                  
&&& $z_c$
   & $d$
      & $3$
         & $d$
            & $d + \lambda$
               & $3$
                  & $4$ \\ 
&&\begin{rotate}{90} dynamic \end{rotate}&&&&&&&\\[-0pt]  
\cline{2-10}
&&&&&&&&&\\[-5pt]
&& 
&$s$
   & N/A
      & N/A
         & 1
            & $\frac{d-2}{d-2+\lambda}$
               & N/A
                  & N/A \\
&&&&&&&&&\\[-5pt]                  
&&& $s_T$
   & N/A
      & $-\frac{d+2}{3}$
         & $\frac{d-2}{d}$
            & $\frac{d-2}{d+\lambda}$
               & $-\frac{d+2}{3}$
                  & $(-\frac{d+4}{4})$$^i$ \\ 
&&\begin{rotate}{90} transport \end{rotate}&&&&&&&\\[-0pt] 
\cline{1-10}\\[-8pt]
\cline{1-10}
&&&&&&&&&\\[-2pt]   
&  & & Widom  & (yes) & yes & yes & yes & yes & yes \\[5pt]
&  & & Essam-Fisher  & yes & yes & yes & yes & yes & yes \\[5pt]
& \begin{rotate}{90} weak \end{rotate}  &  & Fisher   & yes & yes & yes & yes & yes & yes \\[5pt]
\cline{4-10}
&&&&&&&&&\\[-2pt]
&  & \begin{rotate}{90} scaling \end{rotate} & $\alpha$-hyperscaling  & yes & \ \,no$^{\,j}$ & yes & yes & \ \,no$^{\,j}$ & \ \,no$^{\,k}$\\[5pt]
\begin{rotate}{90}  Exponent Relations  \end{rotate} &  & & $\bar\alpha$-hyperscaling  & yes & (yes)$^{\,l}$ & yes & yes & (yes)$^{\,l}$ & (yes)$^{m}$ \\[5pt]
& \begin{rotate}{90} strong \end{rotate} & & $\bar\alpha + 2\beta + \gamma = \nu(z_m + z_c)$$^{\,n}$   & yes & \ \,no$^{\,j}$ & yes & yes & \ \,no$^{\,j}$ & \ \ no$^{\,k}$ \\[5pt]
& & & Wegner  & N/A & no & yes & yes & no & no \\[5pt]
& & & Chayes et al  & N/A & N/A & yes & no$^{\,o}$ & N/A & no \\[5pt]
\hline
\multicolumn{10}{c}{}
\\[-5pt]
\multicolumn{10}{l}{$^{a}$\,For $d>1$.}
\\
\multicolumn{10}{l}{$^{b}$\,This physical fixed point maps onto the unphysical (describing pre-asymptotic behavior only) clean Hertz fixed point. See}
\\
\multicolumn{10}{l}{\ \ \ the text for a discussion of what is observed if the critical point is approached along generic paths in the phase diagram.}
\\
\multicolumn{10}{l}{$^{c}$\,For $d\geq 1$. At the upper critical dimension $d=1$ there are logarithmic corrections to scaling.}
\\
\multicolumn{10}{l}{$^{d}$\,For $2<d<4$. Values in this column equal values in the next column for $\lambda=0$. The critical behavior consists of log-}
\\
\multicolumn{10}{l}{\ \ \ normal terms multiplying power laws with the exponents shown.}
\\
\multicolumn{10}{l}{$^{e}$\,For $2<d<4$. $\lambda$ depends on the distance from criticality. In $d=3$, $\lambda \approx 2/3$ in a large region.}
\\    

\multicolumn{10}{l}{$^{f}$\,For $d\geq 0$. At the upper critical dimension $d=0$ there are logarithmic corrections to scaling.}
\\
\multicolumn{10}{l}{$^{g}$\, For $1\leq d\leq 3$ and $d>3$, respectively.\quad $^{h}$\, For $0<d\leq 4$ and $d>4$, respectively.}
\\
\multicolumn{10}{l}{$^{i}$\, See Sec.~\ref{subsubsec:III.B.1} for the sense in which this result is valid. \quad $^{j}$\, Except for $d=1$. \quad $^{k}$ Except for $d=0$.}
\\
\multicolumn{10}{l}{$^{l}$\, For $1\leq d \leq 3$ only. \quad $^{m}$ For $0 \leq d \leq 4$ only.\quad $^{n}$\, Not an independent relation. \quad $^{o}$ See the text for a discussion.}
\\

\end{tabular}
\end{ruledtabular}
\label{table:1}
\end{table*}

\subsubsection{The first-order transition}
\label{subsubsec:III.A.2}
There are strong theoretical arguments for the quantum phase transition from a paramagnet
to a homogeneous ferromagnet to be first order,\cite{Belitz_Kirkpatrick_Vojta_1999, Kirkpatrick_Belitz_2012b}
and this is indeed the prevalent experimental observation.\cite{Brando_et_al_2015} Here we discuss
the exponent values, and the scaling relations, at this particular first-order quantum phase transition as
an example of the general scaling theory in Sec.~\ref{subsec:II.C}.

Let us first discuss the dynamical exponents. The soft fermionic fluctuations that drive the transition
first order are of a ballistic nature with $z = z_2 = 1$. Their coupling to the order-parameter fluctuations
lead to $z_m = 1$. The static fermionic susceptibility in clean metals at $T=0$ has a wave-number dependence
$\chi({\bm k}\to 0) \propto \text{const.} + \vert k\vert^{d-1}$.\cite{Belitz_Kirkpatrick_Vojta_1997} This plays
against the $\vert\Omega\vert/\vert{\bm k}\vert$ Landau-damping term (see Eq.~(\ref{eq:3.1}) with $a{\bm k}^2$
replaced by $\vert k\vert^{d-1}$), which produces another dynamical exponent $z_1 = d$. For $d>1$ we thus
have $z_c = z_1 = d$. $\beta = \beta_T = 0$ and $\delta = \infty$ by the discontinuous nature of the transition. 
Also, the fact that the order parameter is dimensionless enforces $\nu = 1/(d + z_m) = 1/(d + 1)$ and
$\eta = 2 - d - z_m = 1 - d$, as explained in Sec.~\ref{subsec:II.C}, while $\nu_T = \nu/\nu z_m = 1$. 
It further implies that the free-energy density scales linearly with the control parameter, which implies
$\alpha = 1$ and $\alpha_T = \alpha/\nu z_m = d+1$. Finally, in a fermionic system the specific-heat
coefficient must display a discontinuity at a discontinuous phase transition, which implies $\bar\alpha = \bar\alpha_T = 0$.
Note that, combined with the scaling result (\ref{eq:2.33g}), this implies that the dynamical exponent $z_c$ must
be equal to $d$.

All of these exponent values are displayed in Table~\ref{table:1}. We see that all exponent relations, including those
that rely on strong scaling, are valid as expected. The Widom equation is obviously fulfilled only in the sense that
$\beta(\delta - 1) \to 1$ as $\beta\to 0$, $\delta\to\infty$.

\subsubsection{The quantum wing-critical point}
\label{subsubsec:III.A.3}

\begin{figure}[t]
\includegraphics[width=8cm]{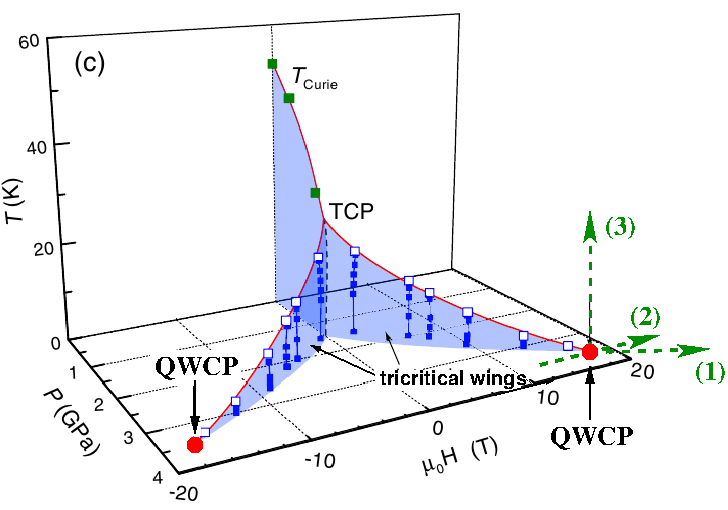}
\caption{Tricritical wings observed in the temperature-pressure-magnetic field phase diagram of UGe$_2$ by Kotegawa et al., 
              Ref.~\onlinecite{Kotegawa_et_al_2011b}. Data points are represented by squares, 
              blue planes are planes of first-order transitions, red solid lines are lines of second-order transitions,
              red points are the quantum wing-critical points (QWCPs). Also shown is the tricritical point (TCP). Dotted
              green arrows labeled (1), (2), (3) are three different paths through a QWCP; the critical behavior
              predicted for each path is discussed in the text. Figure adapted from Ref.~\onlinecite{Kotegawa_et_al_2011b}.}
\label{fig:2}
\end{figure}

Since the quantum ferromagnetic transition is first order, while the corresponding thermal transition is generically
second order, there necessarily is a tricritical point in the phase diagram if the Curie temperature is continuously
suppressed by means of some control parameter. This in turn leads to the existence of tricritical wings, i.e., a pair
of surfaces of first-order transitions that emanate from the coexistence curve in the $h=0$ plane.\cite{Belitz_Kirkpatrick_Rollbuehler_2005} These wings are 
indeed commonly observed, see Fig.\ \ref{fig:2} for an example. They end in a pair of quantum critical points in the 
$T=0$ plane. These quantum wing-critical points (QWCPs; in the literature they are often erroneously referred to as quantum
critical end points) correspond to a fixed point that maps onto Hertz's fixed point; they thus are an example of a
true quantum critical point that is correctly described by Hertz theory, and the critical behavior is known exactly
for all $d>1$.\cite{Belitz_Kirkpatrick_Rollbuehler_2005} 
The critical exponents are thus the same as those discussed in Sec.~\ref{subsubsec:III.A.1}, and the entries in
Table~\ref{table:1} reflect this. However, the application of these results for the prediction of experimental
observations must be handled with care, as we will now discuss.

A crucial point to remember is that at the QWCP, in contrast to Hertz's fixed point, the field $h$ conjugate to the order parameter is not the
physical magnetic field, and $h=0$ only on special paths in the phase diagram that are not natural paths to choose
for an experiment. Let $H$ be the physical magnetic field, and for definiteness let us assume that the control
parameter is hydrostatic pressure $p$. In the three-dimensional $T$-$p$-$H$ phase diagram, let the QWCP be located
at $(0,H_{\text{c}}, p_{\text{c}})$, and let the magnetization at this point have the value $m_{\text{c}}$. Then the
conjugate field is given by $h = 2m_{\text{c}}\delta p - \delta H$, where $\delta p = p - p_{\text{c}}$, and
$\delta H = H - H_{\text{c}}$.\cite{Belitz_Kirkpatrick_Rollbuehler_2005, Brando_et_al_2015} In order to
observe the exponent $\beta$, for instance, one therefore needs to approach the QWCP on a curve given
asymptotically by $\delta H = 2m_{\text{c}} \delta p$, see path (1) in Fig.~\ref{fig:2}. This is in exact analogy to the
case of a classical liquid-gas critical point in the $p\,$-$T$ plane, where an observation of $\beta$ requires that
the critical point be approached on the critical isochore.\cite{Fisher_1983, Chaikin_Lubensky_1995} Probing the
QWCP along a generic path in the $T=0$ plane measures the exponent $\delta$ instead, e.g., 
\be
m(p = p_{\text{c}} ,H, T=0) \propto \vert H - H_{\text{c}}\vert^{1/\delta} =  \vert H - H_{\text{c}}\vert^{1/3}\ .
\label{eq:3.3}
\ee
This is the behavior predicted for the magnetization along path (2) in Fig.~\ref{fig:2}.
A related complication occurs for the temperature dependence of observables at the QWCP. Consider again
the order parameter. The exponent $\beta_T$ is defined via the $T$-dependence of $m$ at criticality in zero
conjugate field, see Appendix~\ref{app:A}. Since $p_{\text{c}}$ is temperature dependent, $\vert p_{\text{c}}(T) - p_{\text{c}}(T=0)\vert \propto T^{(d+1)/3}$,
see Sec.~\ref{subsubsec:III.A.1}, this means that $\beta_T$ can be observed only on a particular surface
in the three-dimensional parameter space spanned by $T$, $p$, and $H$. Along a generic path through the
QWCP the temperature dependence of $m$ is given by the exponent combination $2\nu_T/\delta$. In particular,
just raising the temperature from zero at the critical point (path (3) in Fig.~\ref{fig:2}) yields
\bse
\label{eqs:3.4}
\be
m(p = p_{\text{c}}, H = H_{\text{c}}, T) \propto \vert h\vert^{1/\delta} = - T^{2\nu_T/\delta} = - T^{(d+1)/9}\ .
\label{eq:3.4a}
\ee
This is the result that was derived in Ref.~\onlinecite{Belitz_Kirkpatrick_Rollbuehler_2005}, see also
Ref.~\onlinecite{Brando_et_al_2015}. An observable that is easier to measure is the magnetic susceptibility
$\chi_m$, which along the same path behaves as
\be
\chi_m(p = p_{\text{c}}, H = H_{\text{c}}, T) \propto \vert h\vert^{-\gamma/\beta\delta} = T^{-2(d+1)/9}\ .
\label{eq:3.4b}
\ee
\ese

\subsection{Disordered systems}
\label{subsec:III.B}

Quenched disorder in metallic ferromagnets introduces many different effects. Some, such as the change of the
conduction-electron dynamics from ballistic to diffusive, directly affect the behavior at the quantum phase
transition. Others, such as rare-region effects that can lead to the appearance of a quantum Griffiths region
in the paramagnetic phase,\cite{Vojta_2010, Brando_et_al_2015} are superimposed on critical singularities
and may easily be confused with the latter. Disentangling these various effects is challenging from both a
theoretical and an experimental point of view, and many open questions remain. Here we ignore rare-region
effects and discuss the effects of quenched disorder on the phase transition itself.

This problem was also considered by Hertz,\cite{Hertz_1976} who argued that the only salient change compared
to the clean case is in the
Landau-damping term in Eq.~(\ref{eq:3.1}), which now is $\vert\Omega_n\vert/{\bm k}^2$ due to the diffusive 
dynamics of the conduction electrons. This obviously leads to a dynamical critical exponent $z = 4$, and to 
an upper critical dimension $d_{\text{c}}^+ = 0$. The finite-temperature behavior can be discussed in exact
analogy to Millis's treatment of the clean case in Ref.~\onlinecite{Millis_1993}. 

\begin{figure}[t]
\includegraphics[width=8cm]{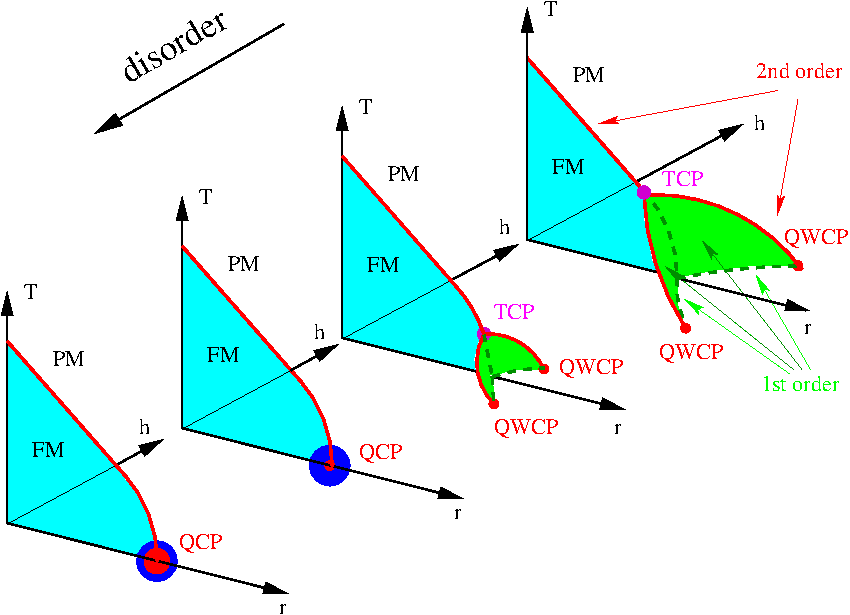}
\caption{Evolution of the phase diagram in temperature-control parameter-field ($T$-$r$-$h$) space with 
              increasing disorder. For weak disorder the phase diagram is
              a schematic version of the experimentally observed one shown in Fig.~\ref{fig:2}. Increasing disorder
             suppresses the tricritical point (TCP). At a critical disorder the TCP merges with the quantum wing-critical
             points (QWCPs), and a quantum critical point (QCP) in zero field emerges. For disorder strengths close
             to the threshold the asymptotic quantum critical behavior is confined to a very small region (red circle),
             and the observable critical behavior (blue circle) is controlled by Hertz's fixed point. With further
             increasing disorder the instability of Hertz's fixed point becomes apparent farther away from the
             transition, and the region controlled by the true critical fixed point grows.}
\label{fig:3}
\end{figure}

Hertz's fixed point in the disordered case is again unstable. The physics behind this instability is the 
same as in the clean case, viz. the coupling of the magnetization to fermionic soft modes, which now are
diffusive in nature at asymptotically small wave numbers. At larger wave numbers they cross over to the
clean soft modes. For weak disorder, this crossover occurs at a very small wave number, and one
expects the effects to be small. This expectation is indeed borne out in practical terms: Even though
strictly speaking there cannot be a first-order transition with any amount of disorder,\cite{Aizenman_Greenblatt_Lebowitz_2012}
the smearing of the transition is very small and the observable effect is just a suppression of the tricritical
temperature, with the quantum phase transition remaining first order.\cite{Sang_Belitz_Kirkpatrick_2014}
However, for a threshold value of the disorder strength the tricritical temperature reaches zero, the quantum
phase transition becomes second order, and above this threshold it remains second order, albeit with an unusual critical behavior. 
The evolution of the phase diagram is shown in Fig.~\ref{fig:3}. A renormalized mean-field theory 
leads to unusual critical exponents, and a renormalization-group analysis of the fluctuations reveals that there 
are marginal operators in all dimensions $2<d<4$, which leads to log-normal terms that multiply the usual critical
power laws.\cite{Kirkpatrick_Belitz_1996, Belitz_et_al_2001a, Belitz_et_al_2001b} This modification of scaling
mimics, in a sizable pre-asymptotic region, power laws with effective exponents that are quite different from the
asymptotic ones.\cite{Kirkpatrick_Belitz_2014} Furthermore, the behavior associated with Hertz's unstable fixed point
can be observable in substantial regimes in parameter space, as we will show below. We therefore list in 
Table~\ref{table:1}, and discuss below, the critical behavior at Hertz's fixed point as well as that associated
with the physical critical fixed point and its pre-asymptotic region.

\subsubsection{Hertz's fixed point}
\label{subsubsec:III.B.1}

As mentioned above, Hertz's fixed point in the disordered case is never a physical fixed point; it is unstable with respect
to the same physical processes as in the clean case, viz., coupling of the magnetization to soft fermionic excitations. 
For disordered systems, the latter are diffusive, and their effects are small in the limit $\kF\ell \gg 1$, with $\kF$ the Fermi
wave number and $\ell$ the elastic mean-free path. In addition, it
is destabilized by a random-mass term in the LGW theory for the magnetic order parameter.\cite{Kirkpatrick_Belitz_1996,
Belitz_et_al_2001a} This effect is small as long as the fluctuations of the distance from criticality, $\delta r$, are small compared to
$r$ itself: $\delta r \propto (\Delta/\xi^d)^{1/2} \propto (\Delta\,r^{d\nu})^{1/2} \ll r$. Here $\Delta$ is the disorder strength, and
we assume that the fluctuation is proportional to the inverse square root of a correlation volume $\xi^d$, with $\xi$ the
correlation length. With mean-field exponents in $d=3$, this condition becomes $\Delta^2 \ll r$. With $\Delta = 1/\kF\ell$ this
is a very weak constraint for small disorder, $\kF\ell \gg 1$, and Hertz's
mean-field description remains valid in a large parameter range. In $d=3$, the condition becomes $\xi\, \ll \ell$.
In Fig.~\ref{fig:3}, the region controlled by Hertz's fixed point is schematically
indicated by the blue region around the quantum critical point, and the asymptotic critical region by the red circle
inside the blue region. We conclude that in many systems one expects sizable regions in the phase
diagram where Hertz's fixed point yields the observable behavior, and only very close to the quantum critical point does
the behavior cross over to the asymptotic one. This experimental relevance of Hertz's fixed point in disordered systems
has not been appreciated before, and we therefore discuss it here. 

Table~\ref{table:1} lists the critical exponents associated with Hertz's fixed point in the presence of disorder. The static
exponents have their mean-field values for all $d>0$, and the dynamical exponent $z_c = 4$ results from the diffusive dynamics
of the conduction electrons, see above. The temperature dependence of the observables, expressed by the $T$-exponents,
is obtained by repeating Millis's analysis with obvious modifications; the most important one being that the scale dimension
of the DIV is now $[u] = -d$, rather than $[u] = -(d-1)$ in the clean case. 

The system is above its upper critical dimensionality for all $d>0$, and the exponent relations that depend on strong scaling
therefore break down, while those that rely only on weak scaling hold, in perfect analogy to the clean case in $d>1$. The
discussion of the exponents $\bar\alpha$ and ${\bar\alpha}_T$ in Sec.~\ref{subsubsec:III.A.1} also carries over with obvious
modifications. At the upper critical dimension $d_c^+ = 0$ hyperscaling holds as expected, and there is only one dynamical
exponent, $z_m = z_{\text{c}} = 4$, as is the case in clean systems in $d=1$. 
An important exponent relation that has no clean analog is the rigorous inequality $\nu > 2/d$.\cite{Chayes_et_al_1986}
This is violated for all $d<4$, which by itself implies that the fixed point cannot be stable. 

We next consider the temperature dependence of the conductivity. The arguments that led to Eq.~(\ref{eq:3.2}) are easily
modified to apply to the disordered case. Since the scale dimension of the DIV is now $[u] = -d$, the effective scale dimension
of $\Delta\sigma$ is $[\Delta\sigma] = -(d+4)$. In addition, the dynamical exponent is now $z_c = 4$, which yields
\be
\Delta\sigma(r,T) = b^{d+4}\,F_{\sigma}(r\,b^{1/\nu}, T\,b^4)\ .
\label{eq:3.5}
\ee
This immediately yields $s_T = -(d+4)/4$ as shown in Table~\ref{table:1}. Here we have assumed that the backscattering
factor is still present, which requires that the residual resistivity is small compared to the temperature-dependent part;
see the discussion below. The exponent $s$ does not exist in this regime for the same reason as in the clean case.

The disorder dependence of the temperature-dependent conductivity can
also be obtained from scaling. Let us start from Eq.~(\ref{eq:3.2}) with $z_c = 3$. Dropping the meaningless $r$-dependence, 
and adding the dependence of $\Delta\sigma$ on the elastic mean-free path $\ell$, we have
\be
\Delta\sigma(T,\ell) = b^{d+2}\,F_{\sigma}(T\,b^3, \ell\,b^{-1})\ .
\label{eq:3.6}
\ee
The scaling function $F_{\sigma}$ now must have the following properties: (1) $F_{\sigma}(x,y\to\infty) = F_{\sigma}^{\text{clean}}(x)$,
so we recover Eq.~(\ref{eq:3.2}). (2) For $\ell/b\to 0$ the prefactor must change from $b^{d+2}$ to $b^{d+4}$, and the temperature
scale factor must change from $b^3$ to $b^4$ in order to recover Eq.~(\ref{eq:3.5}). This is achieved by 
$F_{\sigma}(x,y\to 0) = F_{\sigma}^{\text{disordered}}(x/y)/y^2$. We thus find the following generalization of Eq.~(\ref{eq:3.5}):
\be
\Delta\sigma(T) = b^{d+4}\ell^{-2}\,F_{\sigma}((T/\ell)\,b^{4})\ .
\label{eq:3.7}
\ee
This yields
\be
\Delta\sigma(T) \propto \ell^{-(4-d)/4}\,T^{-(d+4)/4}\ ,
\label{eq:3.8}
\ee
which yields the disorder dependence of the prefactor in addition to the exponent $s_T$. We have checked this result by means
of an explicit calculation based on the Kubo formula for the conductivity, and have found agreement.

The above results for the conductivity depend on various assumptions that limit their regime of validity and require a
discussion. Firstly, we have assumed diffusive electron dynamics, which requires $T\tau < 1$, with $\tau = \ell/v_{\text{F}}$ the
elastic mean-free time. Secondly, we have assumed that the contribution from Eq.~(\ref{eq:3.8}), $\Delta\rho \propto \ell^{(4-d)/4}\,T^{(d+4)/4}$,
which is the contribution from small wave numbers in any transport theory, is larger than the clean contribution, $\Delta\rho \propto T^{(d+2)/3}$.
Thirdly, we have assumed that the backscattering factor is still present. This requires that the residual resistivity
$\rho_0$ be small compared to the temperature-dependent contribution $\rho(T)$. The last two assumptions both imply that the
temperature must be large compared to a disorder-dependent energy scale, and the result (\ref{eq:3.8}) will thus be valid only
in a temperature window. In a simple model for a metallic ferromagnet, where the Fermi energy is the only microscopic energy
scale, this window does not exist, since the last requirement leads to an unrealistically large lower bound for the temperature.
This is misleading, however. In any real ferromagnetic material a complicated band structure leads to multiple microscopic
energy scales, some of which are strongly renormalized downward from the Fermi energy of a nearly-free-electron model.
This is especially true for the low-Curie-temperature ferromagnets that are good candidates for observing a quantum 
ferromagnetic transition. Different factors of the temperature in the above arguments will be normalized by different
microscopic scales, and generically one expects the relevant temperature window to exist. This is certainly true empirically,
as the temperature-dependent resistivity is observed to dominate the residual resistivity for all but the lowest temperatures
in many materials.\cite{Campbell_Fert_1982}

In the regime discussed above the exponent $s$ as defined in Eq.~(\ref{eq:A.6}) does not exist for the same reason as in
the clean case, and the low-temperature behavior of the conductivity away from criticality is again given by Eq.~(\ref{eq:3.2'}).
Remarkably, the $r$-dependence of the prefactor of the $T^{-2}$ dependence is the same in the clean and dirty cases.

\subsubsection{The physical fixed point}
\label{subsubsec:III.B.2}

\begin{figure}[t]
\includegraphics[width=8cm]{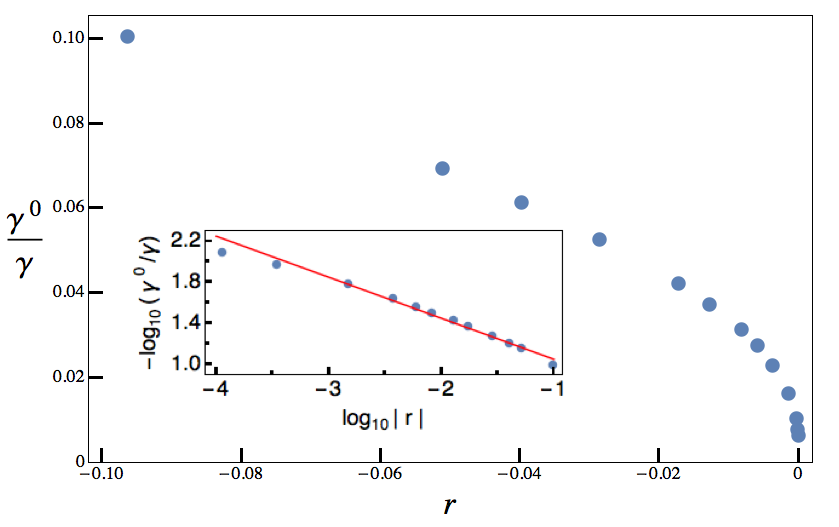}
\caption{The inverse of the specific-heat coefficient $\gamma$ normalized by the Fermi-liquid value $\gamma^0$ as a
              function of $r$ in the ordered phase for $d=3$. The inset demonstrates the effective power law $\gamma \propto \vert r\vert^{-\bar\alpha}$
              with $\bar\alpha = \lambda/(1 + \lambda) \approx 0.4$. Figure adapted from Ref.~\onlinecite{Kirkpatrick_Belitz_2014}.}
\label{fig:4}
\end{figure}

Asymptotically close to the quantum critical point the behavior is described by a Gaussian fixed point with marginal
operators in all dimensions $2<d<4$,\cite{Belitz_et_al_2001a, Belitz_et_al_2001b} and with increasing disorder the region in
parameter space that is controlled by this fixed point grows. The marginal operators result in critical behavior
that is not given by pure power laws. For example, the magnetization at $T=0$ in $d=3$ asymptotically behaves as
\bse
\label{eqs:3.10}
\be
m(r,T=0) \propto (-r)^2\,[g(\ln(-1/r))]^2\ .
\label{eq:3.10a}
\ee
Here the function $g$ is asymptotically log-normal,
\be
g(x\to\infty) \propto e^{\,c\,x^2}\ ,
\label{eq:3.10b}
\ee
\ese
with $c$ a constant; see Ref.~\onlinecite{Belitz_et_al_2001b} for details. In a large pre-asymptotic region these
multiplicative corrections to scaling mimic power laws that span several decades in $r$ or $T$.\cite{Kirkpatrick_Belitz_2014}
For instance, the magnetization obeys an effective homogeneity law
\be
m(r,T,h) = b^{-2}\,F_m(r\,b^{d-2+\lambda}, T\,b^{2+\lambda}, h\,b^{d+\lambda})\ .
\label{eq:3.11}
\ee
$\lambda$ is an effective exponent that depends on $d$ and $r$ and goes to zero for $r\to 0$; however, it is
approximately constant over a large $r$-range. For $d=3$, $\lambda \approx 2/3$ for $0.001 \alt \vert r\vert \alt 0.1$.
As an example, we show in Fig.~\ref{fig:4} the divergence of the specific-heat coefficient, $1/\gamma(r,T=0)~\propto~\vert r\vert^{\bar\alpha}$.
The asymptotic value $\lambda = 0$ needs to be interpreted as signaling the presence of the log-normal terms. 
The resulting power laws are listed in the two middle columns in Table~\ref{table:1}. Since the system is below an
upper critical dimension for $2<d<4$, all exponent relations hold, including the hyperscaling relations. This holds
even for the pre-asymptotic effective exponents, with the exception of the generalized Harris criterion, Eq.~(\ref{eq:C.1}),
which holds only asymptotically.

\section{Summary, and Discussion}
\label{sec:IV}

\subsection{Summary}
\label{subsec:IV.A}

In summary, we have presented two conceptual advances for quantum phase transitions: First, we have analyzed 
relations between critical exponents, also known as scaling relations. Compared to classical transitions, there are
additional exponents to consider that describe the scaling behavior of observables with respect to temperature,
which is independent from the scaling behavior with respect to the control parameter. The exponents describing
the latter (``$r$-exponents'') are related to the ones describing the former (``$T$-exponents'') by means of a dynamical 
critical exponent; however, care must be taken since at many quantum phase
transitions there is more than one dynamical exponent. We have shown that the Widom and Fisher equalities 
hold for both the $r$-exponents and the $T$-exponents under weak assumptions that are fulfilled at most quantum
(and classical) phase transitions. The same is true for the Essam-Fisher equality relating the $r$-exponents. Additional
relations, which are akin to the classical hyperscaling relation between the correlation-length exponent $\nu$ and
the specific-heat exponent $\alpha$, require stronger assumptions that break down if the system is above an
upper critical dimension. We have also generalized the rigorous Rushbrooke inequality to the quantum case,
and we have shown that at quantum phase transitions it is crucial to distinguish between exponents describing
the critical behavior of the specific-heat coefficient, which we denote by ${\bar\alpha}$ and $\bar\alpha_T$, and
the exponents $\alpha$ and $\alpha_T$ that govern the control-parameter susceptibility; at a classical transition,
$\bar\alpha$ and $\alpha$ are the same. We have also discussed Wegner's equality that relates critical exponents
for the electrical conductivity and the correlation-length exponent $\nu$, and have discussed the conditions under
which it is valid. 

Second, we have generalized the concepts of Fisher and Berker related to scaling at classical first-order transitions to
quantum phase transitions. The scaling concepts all carry over, with the temperature playing the role of 
additional dimensions. Again, the presence of multiple time scales leads to complications that need to be
dealt with carefully. With the proper choice of dynamical exponents all scaling relations, including the
hyperscaling ones, hold under very weak assumptions. This reflects the fact that at a first-order transition
there are no dangerous irrelevant variables. 

We then have applied these concepts to the case of the ferromagnetic quantum phase transition as a specific
example. This transition is well suited for this purpose, since it shows very different behavior in the presence
or absence of quenched disorder, respectively, and in either case there are interesting and experimentally
relevant crossover phenomena that are governed by different fixed points, some of which are above and some
of which are at their upper critical dimension in the physically most interesting dimensions $d=2,3$.

\subsection{Discussion}
\label{subsec:IV.B}

We conclude with some discussion points that augment remarks already made in the body of the paper. 

\subsubsection{General points}
\label{subsubsec:IV.B.1}

\paragraph{Number of independent exponents:} 
There are seven $r$-exponents for thermodynamic quantities: 
$\alpha$, $\bar\alpha$, $\beta$, $\gamma$, $\delta$, $\eta$, and $\nu$. These are related by three independent 
weak-scaling exponent relations (Widom, Eq.~(\ref{eq:2.13a}), Essam-Fisher, Eq.~(\ref{eq:2.15a}), and Fisher, 
Eq.~(\ref{eq:2.20a})), and two independent strong-scaling ones ($\alpha$-hyperscaling, Eq.~(\ref{eq:2.23a}), and
$\bar\alpha$ hyperscaling, Eq.~(\ref{eq:2.24a}). In the presence of strong scaling one thus has two independent 
static $r$-exponents. In addition, there are five $T$-exponents: $\alpha_T$, $\bar\alpha_T$, $\beta_T$, and $\gamma_T$ 
that are constrained by two independent weak-scaling exponent relations (Eqs.~(\ref{eq:2.13b}, \ref{eq:2.20b}) plus three
independent strong-scaling ones (Eqs.~(\ref{eq:2.15b}, \ref{eq:2.23b}, \ref{eq:2.24b}). In the presence of strong 
scaling, we thus have two independent static exponents, plus the dynamic ones. In the
absence of strong scaling, there are either four or five independent static exponents plus the dynamic ones,
depending on whether the Essam-Fisher equality for the $T$-exponents holds.

\paragraph{Observability of the exponent $\beta_T$}
The order parameter is nonzero only in the ordered phase, i.e., for $-r > T^{1/\nu z_m}$ (ignoring constant
proportionality factors). For $-r \gg T^{1/\nu z_m}$ one observes static scaling with small temperature
corrections, and for $-r < T^{1/\nu z_m}$ the scaling function vanishes identically. The exponent $\beta_T$
therefore cannot be observed via the $T$-dependence of $m$ at $r=0$. Instead, consider the general weak-scaling
homogeneity law for $m$
\bse
\label{eqs:4.1}
\be
m(r,T) = b^{-\beta/\nu} \Phi_m(r\,b^{1/\nu}, T\,b^{\beta/\nu\beta_T})\ ,
\label{eq:4.1a}
\ee
which follows, e.g., by differentiating Eq.~(\ref{eq:2.6}) with respect to $h$ and using Eqs.~(\ref{eq:2.13a}) and (\ref{eq:2.15a}).
This can be written
\bea
m(r,T) &=& T^{\beta_T}\,\Phi_m(r/T^{\beta_T/\beta},1)
\nonumber\\
          &=& r^{\beta}\,\Phi_m(1,T/r^{\beta/\beta_T})\ ,
\label{eq:4.1b}
\eea
\ese
which defines the exponent $\beta_T$. Note that $\Phi_m(x,1)$ has a zero for some value $x = x_c<0$ that determines
the phase boundary, and that $\Phi_m(x>x_c,1) \equiv 0$. For the temperature derivative of the order parameter this implies
\be
\frac{dm}{dT}(r,t) = T^{\beta_T - 1}\,{\tilde\Phi}_m(r/T^{\beta_T/\beta},1)\ ,
\label{eq:4.2}
\ee
where ${\tilde\Phi}_m(x,y) = \partial_y\,\Phi_m(x,y)$. This has been used to derive Eq.~(\ref{eq:2.22a}) from (\ref{eq:2.21}).

As an illustration, consider the third panel of Fig.~\ref{fig:Ni3AlGa} again. The data imply $m^2 \propto \text{const.} - T^{4/3}$,
where the constant depends on $r$, but the prefactor of the $T^{4/3}$ does not. When interpreted by means of Eq.~(\ref{eq:4.1b})
this implies $\beta = 1/2$ and $\beta_T/\beta = 4/3$. The latter ratio is equal to $1/\nu z_m$, which is consistent with the 
observed shape of the phase boundary in the first panel of the figure. All of this is consistent with the discussion in
Sec.~\ref{subsubsec:III.A.1}. Another way to measure $\beta_T$ is to observe the $T$-dependence of the susceptibility at
criticality, which yields $\gamma_T$, see the second panel in the figure, and use Eq.~(\ref{eq:2.13b}) in conjunction with
an independent determination of the exponent $\delta$. We finally mention that for asymptotically small $m$ at any nonzero
temperature one crosses over to the classical critical behavior characterized by the classical value of the exponent $\beta$.

\paragraph{Coupling of statics and dynamics:} 
Even though the statics and the dynamics are coupled in quantum
statistical mechanics, the Widom, Essam-Fisher, and Fisher exponent equalities do not involve the dynamical
exponents. Even at $T=0$ there thus are exponents relations that do not reflect this coupling.  

\paragraph{Relevance of quenched disorder:} 
Since the generalized Harris criterion by Chayes et al.,
Eq.~(\ref{eq:C.1}), is rigorous it provides a necessary condition for the stability of a critical fixed point in the
presence of quenched disorder. For instance, it immediately tells us that Hertz's fixed point in disordered
systems with its mean-field value $\nu = 1/2$ cannot be stable for $d<4$, although it does not provide any 
hints as to what the fixed point is unstable against, or what replaces it asymptotically. Similarly, it shows
that a first-order transition is strictly speaking impossible in a disordered system, since it would result in 
a correlation-length exponent $\nu = 1/(d + z_m) < 2/d$. 

\paragraph{Significance of $\alpha = 1$ at a first-order QPT:} A first-order classical phase transition is
characterized by the appearance of a latent heat, i.e., a discontinuity of the entropy as a function of the
temperature, and a $\delta$-function contribution to the specific heat. The corresponding physical 
phenomenon at a first-order QPT is a discontinuity of the derivative of the free energy with respect to 
the control parameter, and a $\delta$-function contribution to the second derivative. As an example,
consider a QPT where the control parameter is the hydrostatic pressure $p$. With $g$ the Gibbs free
energy density, the compressibility $\kappa = - \partial^2 g/\partial p^2$ will thus have a $\delta$-function
contribution. That is, at any pressure-driven QPT the system will display a ``latent-volume'' effect, i.e.,
the system volume will change spontaneously and discontinuously at the critical pressure. 

\paragraph{Choice of the control parameter:} 
While deriving a LGW theory, the easiest choice for the
control parameter or mass term $r$ may be a linear combination of some non-thermal parameter and
the temperature, or some power of the temperature. This is not a good choice for $r$ for at least two
reasons: (1) In order to measure, e.g., the exponent $\beta$ it is necessary to keep $T=0$ in addition to
keeping the conjugate field equal to zero. Moving on a path that is not in the $T=0$ plane is akin to
deviating from the critical isochore at a classical liquid-gas transition. (2) The such-obtained temperature
dependence of the phase diagram may not be the leading one due to DIVs. This is what happens, for
instance, in Hertz theory.

\paragraph{Applicability of scaling theory:}

We have restricted ourselves to phase transitions for which a local order parameter exists. We note,
however, that the existence or otherwise of a local order parameter can be a matter of how the
theory is formulated. In the case of the ferromagnetic quantum phase transition in metals that we
have used as an example in Sec.~\ref{sec:III} it is crucial that one treats the fermionic soft modes
that couple to the order parameter explicitly, as in Ref.~\onlinecite{Belitz_et_al_2001a}; if one
integrates them out to formulate a theory entirely in terms of the order parameter a local
description is not possible.\cite{Kirkpatrick_Belitz_1996} We also note that some theories
of ``exotic'' quantum phase transitions that cannot be cast into the language of a simple
Landau theory are structurally very similar to the ferromagnetic quantum phase transition
problem in that they couple an order-parameter field to a gauge field,\cite{Alet_Walczak_Fisher_2006}
or to fermions.\cite{Savary_Moon_Balents_2014}

More generally, we note that the scaling arguments we have employed are extremely
general and hinge only on the existence of a phase transition with power-law critical
behavior. Given this, the free energy will obey a generalized homogeneity law irrespective
of whether or not the transition allows for a traditional Landau description or is of a more
exotic nature. 

\subsubsection{Points related to Hertz's fixed point}
\label{subsubsec:IV.B.2}

\paragraph{Multiple time scales in Hertz theory:} 
We come back to the identification of the dynamical
critical exponent $z_m$ at the beginning of Sec.~\ref{subsubsec:III.A.1}. It is important to distinguish
between theories that truly have multiple time scales that belong to different types of excitations, and
theories where a single time scale may or may not get modified by the effects of DIVs, depending on
the context, which effectively leads to multiple time scales. Hertz's theory, both for the clean and the
disordered case, belongs to the latter category. The time scale with $z_m=1$ in the clean case, or
$z_m = 2$ in the disordered one, at the physical fixed point (see Table~\ref{table:1}) is missing in
Hertz theory, and the effective $z_m$ listed for Hertz's fixed point in Table~\ref{table:1} is due to a DIV 
modifying the sole critical $z=3$ (clean) or $z=4$ (disordered). Note that if a $z=1$ were present
in the clean theory, it would be smaller than the $z_m = 6/(d+1)$ in Hertz-Millis theory, and thus would
play the role of $z_m$ according the arguments in the context of Eqs.~(\ref{eq:2.8}, \ref{eq:2.11}). This
is another indication that Hertz's fixed point cannot be stable.

\paragraph{Numerical values of exponents:} 
The values of the commonly measure exponents 
$\bar{\alpha}_{T}$, $\gamma_{T}$, and $\beta_{T}$ are rather similar at the clean and dirty Hertz fixed 
points, respectively, in $d=3$. The same is true for the scaling of the Curie temperature with the control
parameter, which is $T_{\text{C}} \propto r^{3/4}$ in the clean case, and $T_{\text{C}} \propto r^{4/5}$ in 
the disordered one. This needs to be kept in mind when interpreting experiments.

Also of interest is the value of the conductivity exponent $s_T = -7/4$ at the disordered Hertz fixed
point in $d=3$. This result is a good candidate for explaining the fact that the resistivity is commonly
observed to vary as $T^x$ with $x > 1$ near a ferromagnetic quantum critical point, as the crossover
to the asymptotic critical behavior is expected to occur only at very low temperatures.

\subsubsection{Points related to the physical quantum ferromagnetic fixed point}
\label{subsubsec:IV.B.3}

\paragraph{Significance of $z_c = d$:}
In Sec.~\ref{subsubsec:III.A.2} we showed that $z_c = d$ follows from the first-order nature of the
quantum phase transition. Combined with the origin of the dynamical exponent in a LGW theory,
where $z_c$ arises from a combination of the Landau-damping term with the spin susceptibility
in the paramagnetic phase, this implies that in a clean Fermi liquid the leading non-analytic
wave-number dependence must scale as $k^{d-1}$. The connection between these seemingly
unrelated results is scaling, see Sec.~\ref{subsubsec:III.A.2} and Ref.~\onlinecite{Belitz_Kirkpatrick_2014}.

In this context we also note that at any first-order quantum phase transition in a fermionic system one expects the
specific-heat coefficient to be discontinuous. This implies $\bar\alpha = {\bar\alpha}_T = 0$,
which in turn requires $z_c = d$, see Eq.~(\ref{eq:2.33g}).

\paragraph{Scale dimension of the conjugate field:} 
We saw in Sec.~\ref{subsec:II.C} that a
dimensionless order parameter implies a scale dimension $[h] = d + z_m$ for the conjugate field.
This illustrates the fact that in general it is not possible to simply relate the scaling of $h$ to the scaling
of the energy or temperature. For instance, one might argue that the scale dimension of $h$ should be
equal to $z_c$, since $h$ determines the Zeeman energy. While this is sometimes true (for instance,
at the disordered critical fixed point), it is not true in general. In this context we also note that the
Rushbrooke inequalities, Eqs.~(\ref{eqs:2.22}), do not depend on $[h]$, as $[h]$ cancels between
the contributions $2\beta_T$ and $\gamma_T$ (or $2\beta$ and $\gamma$), respectively, on the
left-hand side.

\paragraph{Logarithmic corrections to scaling in a range of dimensions:} 
The theory of Refs.~\onlinecite{Belitz_et_al_2001a, Belitz_et_al_2001b} yields logarithmic corrections
to scaling (in the sense of log-normal terms multiplying power laws) in an entire range of dimensions,
$2<d<4$. This is unusual; more commonly logarithmic corrections to scaling occur only in a specific
dimension. This can be understood within Wegner's\cite{Wegner_1976b} classification of logarithmic
corrections, as has been discussed in Ref.~\onlinecite{Belitz_et_al_2001b}. The dynamical critical
exponent $z_c = d$ leads to operators that are marginal in a range of dimensions, and this in turn
causes the logarithmic terms. We also note that the mathematical problem of determining the
critical behavior for the theory of Refs.~\onlinecite{Belitz_et_al_2001a, Belitz_et_al_2001b} was
solved exactly in Ref.~\onlinecite{Kirkpatrick_Belitz_1992b}, although the physical interpretation
was unclear at that time. See also Ref.~\onlinecite{Belitz_Kirkpatrick_1994}.

\paragraph{Scaling of the Gr{\"u}neisen parameter}
The Gr{\"u}neisen parameter $\Gamma = (-\partial s/\partial p)/T(\partial s/\partial T)$, with $s$ the
entropy density, is defined as the ratio of the thermal expansion coefficient
$(\partial V/\partial T)_{p,N}/V = -(\partial s/\partial p)_{T,N}$ and the specific heat
$c = T\partial s/\partial T$. It was shown in Ref.\ \onlinecite{Zhu_et_al_2003} that at a pressure-tuned
QPT with a single dynamical exponent $z$ the Gr{\"u}neisen parameter diverges as $\Gamma \propto T^{-1/\nu z}$;
this is readily confirmed by using Eq.~(\ref{eq:2.5}). With two dynamical exponents $z_1>z_2$ we find from
Eq.~(\ref{eq:2.8}) $\Gamma \propto T^{-1/\nu z_1}$. In particular, at the physical fixed point in disordered
metallic ferromagnets we have $\Gamma \propto T^{-1/3}$ in $d=3$. Another useful observation is that,
according to its definition, $\Gamma$ scales as $\Gamma \sim 1/p \sim 1/r$ at a pressure-tuned QPT.
Since $r \sim T^{1/\nu z}$ at a critical point, this is equivalent to the result given above. At the first-order
transition in clean systems, $\Gamma \sim 1/r$ reflects the same $\delta$-function contribution that was
discussed for the compressibility in Sec.~\ref{subsubsec:IV.B.1}. This is also apparent from the fact
that $s \sim \gamma T$, and hence $\Gamma \sim \partial\gamma/\partial p$, and the specific-heat
coefficient $\gamma$ has a discontinuity across the first-order transition.

The behavior at the clean Hertz fixed point has been discussed in Ref.~\onlinecite{Zhu_et_al_2003}. Up
to logarithmic corrections, $\Gamma \sim r^{-1} \sim T^{-2/3}$ in agreement with the above arguments
and $\nu = 1/2$, $z_c = 3$ from Table~\ref{table:1}. At the disordered Hertz fixed point the corresponding
result is $\Gamma \sim r^{-1} \sim T^{-1/2}$.

\acknowledgments
This work was supported by the NSF under Grants No. DMR-1401410 and No. DMR-1401449.

\appendix

\section{Definitions of critical exponents}
\label{app:A}

Let $T$ be the temperature, $h$ the field conjugate to the order parameter, $r$ the control parameter, i.e., the dimensionless distance from criticality at $T=h=0$, 
and $f$ a suitable free-energy density. Consider the correlation length $\xi$, the order parameter $m = \partial f/\partial h$,
the order-parameter susceptibility $\chi_m = \partial^2 f/\partial h^2$, and the specific-heat coefficient $\gamma = c/T = -\partial^2 f/\partial T^2$ as functions 
of $r$, $T$, and $h$, and the susceptibility also as a function of the wave number $k$. Further consider the susceptibility $\chi_r = -\partial^2 f/\partial r^2$,
which we refer to as the control-parameter susceptibility. (More precisely, $\chi_r$ is the susceptibility of the thermodynamic quantity whose conjugate field
is the control parameter.) We define critical exponents at a quantum phase transition as follows.
\medskip\par\noindent
{\it Correlation length:} 
\be
\xi(r\to 0,T=0) \propto \vert r\vert^{-\nu}\quad,\quad \xi(r=0,T\to 0) \propto T^{-\nu_T}\ .
\label{eq:A.1}
\ee
\par\noindent
{\it Order parameter:}
\bea
m(r\to 0,T=0,h=0) &\propto& (-r)^{\beta}\ ,
\nonumber\\
m(r=0,T=0,h\to 0) &\propto& h^{1/\delta}\ ,
\nonumber\\
m(r = 0,T\to 0,h=0) &\propto& T^{\beta_T}\ .
\label{eq:A.2}
\eea
The last definition is purely formal: Since the magnetization is nonzero only in the ordered phase, the $T^{\beta_T}$ in the last line
has a zero prefactor. See the discussion in Sec.~\ref{subsubsec:IV.B.1} of how to interpret the exponent $\beta_T$.
\medskip\par\noindent
{\it Order-parameter susceptibility:}
\bea
\chi_m(r\to 0,T=0;k=0) &\propto& \vert r\vert^{-\gamma}\ ,
\nonumber\\
\chi_m(r=0,T\to 0;k=0) &\propto& T^{-\gamma_T}\ ,
\nonumber\\
\chi_m(r=0,T=0,k\to 0) &\propto& 1/k^{2-\eta}\ .
\label{eq:A.3}
\eea
\par\noindent
{\it Specific-heat coefficient:}
\be
\gamma(r\to 0,T=0) \propto \vert r\vert^{-{\bar\alpha}}\quad,\quad \gamma(r=0,T\to 0) \propto T^{-{\bar\alpha}_T}\ .
\label{eq:A.4}
\ee
\par\noindent
{\it Control-parameter susceptiblity}
\be
\chi_r(r\to 0, T=0) \propto \vert r\vert^{-\alpha}\ ,\ \chi_r(r=0,T\to 0) \propto T^{-\alpha_T}\ .
\label{eq:A.5}
\ee

$\nu$, $\beta$, $\gamma$, $\delta$, and $\eta$ are defined in analogy to the corresponding exponents
at a classical phase transition.\cite{Stanley_1971} In the main text we refer to these exponents, and also to $\bar\alpha$ and $\alpha$, as the
$r$-exponents. The definition of ${\bar\alpha}$ deviates from the one of the classical
exponent customarily denoted by $\alpha$, which is defined in terms of the specific heat rather than
the specific-heat coefficient.
This is necessary in order to factor out the factor of $T$ in the relation between the specific heat
and the specific-heat coefficient,\cite{gamma_footnote} which makes no difference at a thermal phase transition, but goes to
zero at a QCP. For instance, the thermodynamic identity that underlies the Rushbrooke inequality,
Eq.\ (\ref{eq:B.5}), has no explicit $T$-dependence only if it is formulated in terms of specific-heat
coefficients rather than the specific heats. At a classical phase transition, $\bar\alpha$ 
coincides with $\alpha$. Our definition of $\alpha$ at a QPT is analogous to the definition of the
classical exponent $\alpha$ from a scaling point of view. However, the physical interpretation of
$\alpha$ and the susceptibility $\chi_r$ depends on the nature of the control parameter. For instance, 
if the control parameter is hydrostatic pressure, and $f$
is the Gibbs free energy density, then $\chi_r$ is proportional to the compressibility of the system.
At a thermal transition, where $r \propto T - \Tc$, $\chi_r$ is identical with the specific-heat coefficient $\gamma$,
 and $\alpha$ has its usual meaning.
$\alpha_T$, ${\bar\alpha}_T$, $\nu_T$, $\beta_T$, and $\gamma_T$ reflect the fact that a QPT can
be approached either in the $T=0$ plane, or from $T>0$. These exponents are referred to as the
$T$-exponents in the main text. 

We finally define exponents $s$ and $s_T$ that describe the behavior of the electrical conductivity $\sigma$
at the QPT. With $\Delta\sigma$ the scaling part of $\sigma$, these are defined as follows:
\medskip\par\noindent
{\it Electrical conductivity:}
\be
\Delta\sigma(r\to 0, T=0) \propto \vert r\vert^s\quad,\quad \Delta\sigma(r=0,T\to 0) \propto T^{s_T}\ .
\label{eq:A.6}
\ee
The exponent $s$ obviously makes sense only in systems with quenched disorder.

\section{Classical scaling relations}
\label{app:B}

For the convenience of the reader we recall some well-known classical exponent relations (see, e.g., Refs.~\onlinecite{Stanley_1971, Fisher_1983}): 
\bse
\label{eqs:B.1}
\bea
\gamma &=& \beta (\delta - 1)\qquad \text{(Widom)}
\label{eq:B.1a}\\
{\alpha} + 2\beta + \gamma &=& 2 \qquad\qquad \text{(Essam-Fisher)}
\label{eq:B.1b}\\
\gamma &=& (2 - \eta)\,\nu \qquad \text{(Fisher)}
\label{eq:B.1c}\\
\nu\,d &=& 2 - {\alpha} \qquad \text{(``hyperscaling'')}
\label{eq:B.1d}
\eea
\ese
where $\alpha$ is the ordinary specific-heat exponent. The first two equalities follow from a
weak scaling assumption, in the sense of Sec.~\ref{subsec:II.A}, for the singular part $f$ of the free-energy density. This has the
effects of dangerous irrelevant variables, if any, built in and can be written, for instance, as
\be
f(r,h) = b^{-(2 - \alpha)/\nu}\,\Phi_f(r\,b^{1/\nu}, h\,b^{\beta\delta/\nu})\ ,
\label{eq:B.2}
\ee
with $\Phi_f$ a scaling function.
The Fisher scaling relation, Eq.~(\ref{eq:B.1c}), requires an additional (weak) scaling assumption,\cite{Fisher_1983} 
namely, that scaling works for correlation functions as well as for thermodynamic quantities. If we use a
homogeneity law for the wave-number dependent order-parameter susceptibility,
\be
\chi(r,k) = b^{\gamma/\nu}\,\Phi_{\chi}(r\,b^{1/\nu},k\,b)\ ,
\label{eq:B.3}
\ee
and put $r=0$ we obtain Eq.~(\ref{eq:B.1c}).
The hyperscaling relation requires yet another assumption,\cite{Fisher_1983} which is a strong scaling assumption
in the sense of Sec.~\ref{subsec:II.A} and which is equivalent to saying that the scale
dimension of the free energy density is equal to $d$:
\be
f(r,h) = b^{-d}\,\Phi_f(r\,b^{1/\nu}, h\,b^{\beta\delta/\nu})\ .
\label{eq:B.4}
\ee
If hyperscaling holds, the six exponents $\alpha$, $\beta$, $\gamma$, $\delta$, $\eta$, and $\nu$ are
constrained by the four relations in Eqs.~(\ref{eqs:B.1}), and only two of the exponents are independent.

We also list the Rushbrooke inequality
\be
\alpha + 2\beta + \gamma \geq 2\ ,
\label{eq:B.5}
\ee
which holds rigorously, as it only depends on thermodynamic stability conditions. There
are other inequalities that depend on various assumptions.\cite{Stanley_1971}

\section{The Harris criterion}
\label{app:C}

Harris considered the effects of quenched disorder on classical critical points. He found that
the fixed point describing the transition in the clean system is stable with respect to disorder, i.e., the critical behavior is
unaffected by the disorder, as long as the specific-heat exponent is negative, $\alpha < 0$.\cite{Harris_1974}
This statement is referred to as the Harris criterion. 
Chayes et al\cite{Chayes_et_al_1986} generalized Harris's argument and made it rigorous to show that, under very
general conditions, the correlation-length exponent at a critical point in a disordered system,
classical or quantum, must obey
\be
\nu \geq 2/d \qquad \text{(Chayes-Chayes-Fisher-Spencer)}\ .
\label{eq:C.1}
\ee


\begin{thebibliography}{59}
\bibitem[{OP_()}]{OP_footnote}
\bibinfo{note}{Throughout this paper we restrict ourselves to phase transitions
  for which a local order parameter exists. See, however, the last paragraph in
  Sec.~\ref{subsubsec:IV.B.1}}.
\bibitem[{cri()}]{critical_isochore_footnote}
\bibinfo{note}{For a ferromagnetic transition, this means approaching the Curie
  temperature in zero external field; for the liquid-gas transition, it means
  approaching the critical point along the critical isochore.}
\bibitem[{exp()}]{exponents_prime_footnote}
\bibinfo{note}{Throughout this paper we will assume that exponents that are
  defined on either side of the phase transition have the same value on either
  side, as is almost always the case in real systems}.
\bibitem[{sca()}]{scaling_parts_footnote}
\bibinfo{note}{Here, and in all other homogeneity laws, it is tacitly
  understood that it is only the scaling part of the observable in question
  that obeys the homogeneity law. In general, there also is a non-scaling part
  that shows no singular behavior at the phase transition}.
\bibitem[{ent()}]{entropy_scaling_footnote}
\bibinfo{note}{This argument hinges on strong scaling. DIVs can in principle
  invalidate it, but in that case the system is above its upper critical
  dimensionality and the problem can usually be solved exactly.}
\bibitem[{gam()}]{gamma_footnote}
\bibinfo{note}{Here we assume that far from the transition the specific heat
  for $T\to 0$ is given by $c = \gamma\,T$ as is the case, for instance, in a
  Fermi liquid. If one deals with a system where $c = {\tilde\gamma}\,T^a$,
  with $a\neq 1$, then one should formulate the identity expressed by
  Eq.~(\ref{eq:2.21}) in terms of $\tilde\gamma$ instead of $\gamma$.}
\bibitem[{con()}]{conductivity_scaling_footnote}
\bibinfo{note}{We also note that the strong-scaling argument that leads to
  Eq.~(\ref{eq:2.28}) works for the conductivity, but not for the resistivity
  if a non-scaling part is present. This is related to the fact that only the
  conductivity can be written in terms of a time-correlation function via a
  Kubo formula.}
\bibitem[{Lan()}]{Landau_expansion_footnote}
\bibinfo{note}{This argument, while useful, is of a purely formal nature since
  a dimensionless OP means that a Landau expansion in powers of the OP is not
  valid.}
\bibitem[{ucd()}]{ucd_footnote}
\bibinfo{note}{This statement assumes that the action in the form of
  Eq.~(\ref{eq:3.1}) remains valid in $d=1$. This is not the case, since the
  Landau-damping term $\vert\Omega_n\vert/\vert{\bm k}\vert$ reflects the
  properties of a Fermi liquid, which does not exist in $d=1$. Nevertheless, it
  is illustrative to consider Hertz's action as a model that can be analyzed in
  any spatial dimension, which is the attitude we are taking here. For a model
  that treats itinerant ferromagnets in $d=1$ proper, see
  Ref.~\onlinecite{Yang_2004}.}
\end{thebibliography}

\end{document}